\documentclass[iop]{emulateapj}

\shorttitle{Accretion of supersonic winds onto black holes in 3D}
\shortauthors{Gracia-Linares \& Guzm\'an}

\begin{document}

\title{Accretion of supersonic winds onto black holes in 3D: stability of the shock cone}


\author{M. Gracia-Linares and F. S. Guzm\'an}
\affiliation{Instituto de F\'{\i}sica y Matem\'{a}ticas, Universidad
              Michoacana de San Nicol\'as de Hidalgo. Edificio C-3, Cd.
              Universitaria, 58040 Morelia, Michoac\'{a}n,
              M\'{e}xico.}

\begin{abstract}
Using numerical simulations we present the accretion of supersonic winds onto a rotating black hole in three dimensions. We study five representative directions of the wind with respect to the axis of rotation of the black hole and focus on the evolution and stability of the high density shock cone that is formed during the process. We explore both, the regime in which the shock cone is expected to be stable in order to confirm previous results obtained with two dimensional simulations, and the regime in which the shock cone is expected to show a flip-flop type of instability. The methods used to attempt triggering the instability were first the accumulation of numerical errors and second the explicit application of a perturbation on the velocity field after the shock-cone was formed. The result is negative, that is, we did not find the flip-flop instability within the parameter space we explored, which includes cases that are expected to be unstable.
\end{abstract}




\section{Introduction}

The Bondi-Hoyle-Littleton or wind accretion process occurs when a compact object moves with respect to an uniformly distributed perfect fluid considered free of self gravity \citep{Bondi1944,Bondi}. 
This problem has been studied analytically and numerically in Newtonian \citep{frixell1987,matsuda1987,shima1988,sawada1989,matsuda1992,matsuda1991} and relativistic regimes \citep{petrich1988,Font1998a,donmez,cruz2012,Lora2013}. 
In the relativistic regime this process has been associated to high energy sources, for a review see \citep{edgar2004}. The process has been analyzed for various types of compact objects acting as accretors and in this paper we concentrate on the case of a black hole. In its most advanced fashion, the BHL accretion on rotating black holes includes the radiation of the fluid restricted to 2D \citep{zannoti2011,park2013}, 2.5D magnetized fluids in \citep{penner2010,takahashi} and  the case when the fluid has density gradients \citep{LoraCruz2015}.

One of the important aspects of this accretion process is that when the relative speed between the accretor and the fluid is supersonic, a high density region with a conic shape, called shock cone is formed on the rare part of the accretor. When the compact object is rotating, the fluid of the shock cone gets dragged by rotation, eventually distorted and its motion may be unstable (see e.g. \citep{foglizzo2005}). The stability is important because it would provide a model for abrupt acceleration of gas in high density regions near a compact object, which in turn could help modeling short and high energy emissions \citep{donmez,Lora2013}. 

The stability of the shock cone is described in terms of whether or not the flow is steady near the accretor. Particularly interesting is the instability of the flip-flop (FF) type, which happens when the angular momentum of the shock cone starts changing sign and eventually the cone blows up. This instability involves on the one hand the nature of the accretor, for instance a black hole with an event horizon and a neutron star may show different behavior due to the different nature of their surfaces, on the other hand the stability depends on the dynamical conditions of the wind, for instance its velocity and the speed of sound associated to its equation of state.

Empirical criteria have been proposed to establish a regime of stability of the shock cone, independent of the accretor  and more focused on the relation of the accretor size and Bondi accretion radius (e.g. \citep{foglizzo2005}). Given that this analysis requires 
the evolution of the wind by numerical means, for example, on top of the curved background space-time of a black hole, there is still controversy about whether the instability is physical or due to the numerical methods used. For instance, in \citep{shima1988,sawada1989,matsuda1991,matsuda1992,foglizzo2005} some dynamical arguments are found that show that the instability is physical, whereas in \citep{frixell1987,shima1998,cruz2012} it is shown that it might be due to an inadequate numerical implementation.

In this paper we focus on the FF instability of the shock cone for the particular case in which the accretor is a spinning black hole and the wind is a relativistic fluid obeying an ideal gas equation of state in 3D. This problem has been analyzed in 2D in \citep{Font1998a,donmez,cruz2012} where a relativistic study of nonspherical supersonic Bondi-Hoyle accretion onto a spinning black hole with relative accretor size $r_{hor}/r_{acc}=0.1,0.26$ was made on the equatorial plane assuming s-lab symmetry, where $r_{hor} $ is an approximate black hole radius and $r_{acc}$ is the Bondi accretion radius. We present these cases in full 3D because, as far as we can tell, such results have never been presented so far.

According to \citep{foglizzo2005}, the instability is expected to happen when the properties of the system involve a supersonic wind velocity with Mach numbers ${\cal M}^R_{\infty} \ge 2.4$ and the size of the accretor with respect to the Bondi accretion radius is such that $r_{hor}/r_{acc} \alt 0.05$. Therefore, in order to confirm or discard the stability in this regime, we explore the parameter space with this high speed and small enough black holes that lie within this regime, where $r_{hor}/r_{acc}$ ranges from 0.011 to 0.025. We explore the accretion onto  black holes with angular momentum parameter  as high as $a=0.95$ and five representative orientations of the wind flow with respect to the axis of rotation of the black hole. In all the cases analyzed we found that there is no FF instability.

The paper is organized as follows. In section \ref{sec:numerical_set_up} we describe the RHD equations modeling the fluid and the numerical methods used. In section \ref{sec:results} we present the set of configurations we experiment with and finally in \ref{sec:conclusions} we describe the conclusions from our analysis.

\section{Numerical set up}
\label{sec:numerical_set_up}

\subsection{The equations and numerical methods}

The equations of general relativistic hydrodynamics are written as a flux conservative system, that assumes a standard 3+1 decomposition of the space-time, known as the  Valencia formulation \citep{banyuls1997}:

\begin{equation}
  \label{eq:conservative}
  \partial_t {\bf u} + \partial_{x^i} {\bf F}^{(i)} ({\bf u}) = {\bf
    S} ({\bf u}),
\end{equation}

\noindent where ${\bf u}=\{D,S_i,\tau\}$ is the set of conserved variables, $D$ is the generalized rest mass density of the fluid, $S_i$ are the generalized momenta in each direction, $\tau$ is the internal energy, ${\bf F}^{(i)} ({\bf u})$ the fluxes and ${\bf S} ({\bf u})$ the source. In terms of the primitive variables of the fluid elements, $\rho$ the rest-mass density, $p$ the pressure, $v^i$ the fluid 3-velocity, $\epsilon$ the specific internal energy, the conserved variables are defined by

\begin{eqnarray}
  \label{eq:prim2con}
   D &=& \sqrt{\gamma}W\rho \nonumber \\
   S_i &=& \sqrt{\gamma} \rho h W^2 v_i \\
   \tau &=& \sqrt{\gamma}\left( \rho h W^2 - p\right) - D,\nonumber  
\end{eqnarray}

\noindent where $\gamma$ is the determinant of the spatial 3-metric $\gamma_{ij}$ of the three-dimensional spatial slices the space-time is foliated with, $h \equiv 1 + \epsilon + p/\rho$ is the specific entalphy and $W=(1-\gamma_{ij}v^i v^j)^{-1/2}$ is the Lorentz factor. As usual, the system of equations is closed with an equation of state. In terms of primitive and conservative variables the fluxes and sources are:

\begin{eqnarray}
        \  {\mathbf F}^i({\bf u}) &=& \left(
	      \begin{array}{c}
			(\alpha v^i - \beta^i)D\\
	        (\alpha v^i-\beta^i)S_j + p\delta^i_j\\
	         (\alpha v^i - \beta^i)\tau + pv^i
	      \end{array} \right),\nonumber\\
        \  \mathbf{S({\bf u})} &=& \left(
	      \begin{array}{c}
			0\\
	        T^{\mu\nu}\big (\partial_\mu g_{\nu j} +\Gamma^\delta_{\mu\nu} g_{\delta j}\big )\\
	        \alpha \big (T^{\mu0}\partial_\mu \ln \alpha - T^{\mu\nu}\Gamma^0_{\nu\mu} \big)
	      \end{array} \right),\nonumber
\end{eqnarray}

\noindent where $g_{\mu\nu}$ are the components of the 4-metric and $\Gamma^{\delta}{}_{\mu\nu}$ the Christoffel symbols of the space-time, $\alpha$ the lapse function and $\beta^{i}$ the components of the shift vector associated to the 3+1 decomposition of the space-time.

We solve the system of equations (\ref{eq:conservative}) using the Cactus Einstein Toolkit (ETK)  \citep{ETK2012,ETK} which provides the necessary computational tools to evolve this relativistic fluid in full 3D. In particular we use the GRHydro Thorn \citep{baiotti2005}, which contains  high resolution shock capturing methods to solve these equations. For these methods we specifically use the HLLE and Marquina numerical flux formulae and minmod and MC linear reconstructors. For the integration in time we use a fourth order Runge-Kutta method. As part of the diagnostics tools, we use the Outflow Thorn that allows to measure the flux of rest mass across a given spherical surface and monitor the behavior of the accretion rate.

\subsection{Initial conditions}

We describe the space-time of the rotating black hole using Kerr-Schild (KS) horizon penetrating coordinates, with the axis of rotation parallel to $\hat{z}$ as follows:

\begin{eqnarray}
ds^2 &=& \left(\eta_{\mu\nu} + \frac{2Mr^3}{r^4+a^2z^2}l_{\mu}l_{\nu}\right)dx^{\mu}dx^{\nu},\nonumber\\
l_{\mu} &=& \left(1,\frac{rx+ay}{r^2+a^2},\frac{ry-ax}{r^2+a^2},\frac{z}{r}\right),\nonumber\\
r &=& \sqrt{\frac{r_*^{2}-a^2+\sqrt{(r_*^{2}-a^2)^2+4a^2z^2}}{2}},\nonumber\\
r_{*} &=& \sqrt{x^2 + y^2 + z^2},
\end{eqnarray}

\noindent where $M$ is the mass of the black hole and we have assumed geometric units $G=c=1$ to hold.

The fluid is set initially as a spatially constant rest mass density ideal gas, moving toward the black hole and  with a given asymptotic velocity $v_{\infty}^2=v^iv_i$. We assume the fluid obeys a gamma-law equation of state $p=(\Gamma-1)\rho \epsilon$ and thus we introduce the asymptotic speed of sound $c_{s \infty}$. Once we define the value of $c_{s\infty}$ and assume the density to be initially a constant $\rho = \rho_{\mathrm{ini}}$, the pressure can be written as $p_{\mathrm{ini}} = c_{\mathrm{s} \infty}^2 \rho_{\mathrm{ini}}/(\Gamma - c_{\mathrm{s} \infty}^2 \Gamma_1)$, where $\Gamma_1=\Gamma/(\Gamma -1 )$. In order to avoid negative and zero values of the pressure, the condition $c_{\mathrm{s}\infty} < \sqrt{\Gamma - 1}$ has to be satisfied. Finally, with this value for $p_{\mathrm{ini}}$, the initial internal specific energy is reconstructed using the equation of state. 

In order to parametrize the initial data, we define the relativistic Mach number at infinity ${\cal M}^R_{\infty} = W v_{\infty} /W_s c_{\mathrm{s}\infty}$, where $W$ is the Lorentz factor of the gas and $W_s$ is the Lorentz factor calculated with the speed of sound. When ${\cal M}^R_{\infty}$ is bigger than one, the wind is said to be supersonic and otherwise subsonic. 

According to \citep{foglizzo2005}, the two parameters deciding on the stability of the process are the Mach number of the gas ${\cal M}^R_{\infty}$ and the relative size of the accretor with respect to the Bondi accretion radius $r_{acc}$. In our case the compact object is a black hole and we define an approximate black hole radius to be $r_{hor}=2M$, even though our hole is rotating. Thus the relative size of the accretor is given by $r_{hor}/r_{acc}=v_{\infty}^2+c^2_{\mathrm{s}\infty}$. 

This formula indicates that a relatively small black hole is equivalent to have the case of a wind moving with a small velocity $v_{\infty}$. This explains why the cases where $r_{hor}/r_{acc}$ is small are difficult to track, and even more difficult in 3D. The reason is that the numerical domain is set such that it contains at least a sphere of radius $r_{acc}$ in order to see the accretion process. Slower winds imply a bigger $r_{acc}$, therefore demand a bigger numerical domain and consequently more computational resources.

In order to observe the behavior of the process in a general scenario, we choose five representative different directions of the wind, that we characterize in terms of its orientation with respect to $\hat{z}$ which is the axis of rotation of the black hole denoted by $\Uparrow$. These five different orientations of the wind are represented by $\uparrow~ \downarrow~ \rightarrow~ \swarrow ~\nwarrow$, and correspond to directions parallel to $\hat{z},~ -\hat{z},~\hat{x},~-\hat{x}-\hat{z}$ and $-\hat{x}+\hat{z}$ respectively. 

\subsection{Boundary conditions}

The ETK allows the solution of the equations in Cartesian coordinates, and thus the domain is a cubic box of a given size. In the case of the evolution of a wind, two boundary conditions are required (see e.g. \citep{donmez,cruz2012}), namely an inflow boundary condition in the upstream region of the domain, where the gas is being pumped into the box, and outflow boundary conditions in the piece of boundary where the wind is expected to exit the numerical domain.

The ETK does not provide the tools to automatically incorporate different boundary conditions on different faces of the cubic domain and we programmed a module that allows one to do it.

There is another very important boundary to consider. Since we are using horizon penetrating coordinates we are allowed to perform an excision inside the black hole horizon as done in \citep{cruz2012,Lora2013}. The excision method consists in the remotion of a chunk of the numerical domain inside the black hole horizon.  The removed chunk is usually assumed to be a lego sphere which defines a lego shaped boundary inside the horizon. Considering the light cones in KS coordinates remain open and point toward the singularity inside the black hole horizon, no boundary conditions are needed, and extrapolation of the fluid variables suffices to guarantee that the fluid is not reflected back toward the exterior of the black hole. This method is already implemented in the ETK as described in \citep{hawke2005}.

\section{Numerical experiments and results}
\label{sec:results}

Showing the instability of a given fluid configuration requires the perturbation theory analysis, which applies when equilibrium configurations are given a priori. Another possibility uses the numerical evolution of equilibrium configurations and then numerical errors, for instance truncation or discretization errors act as the perturbation and eventually (or not) trigger an instability. When the configuration is stable, the perturbation is simply expelled and the system remains in a state pretty much the unperturbed one, whereas in unstable cases the variables of the system change notably.

In the present case we deal with the stability of a shock cone, which is a configuration that actually forms as a consequence of the evolution of the fluid and there is no equilibrium configuration given a priori. Thus one needs first to evolve the system until the shock cone forms. The shape and properties of the cone are not constructed based on a set of assumptions on the density profile or velocity field, but merely given by numerical evolution. Therefore, in order to know about its stability, the method consists in tracking the further evolution of the system, we simply need to evolve the system during a sufficiently long time and wait for the numerical errors to cumulate and then possibly make the cone to oscillate in a FF fashion or show any other type of instability. 

The different configurations of a wind accretion process are characterized by the following parameters: 1) the asymptotic velocity of the wind, 2) the initial rest mass density of the wind, 3) the adiabatic index $\Gamma$, 4) the black hole angular momentum $a$, 5) the orientation of the wind with respect to the axis of rotation of the black hole and 6) the relative accretor size $r_{hor}/r_{acc}$.

According to \citep{foglizzo2005}, the size of the accretor and the velocity of the wind are the parameters that determine whether the shock cone is stable or unstable, and it has been estimated that in the range of relative accretor size $r_{hor}/r_{acc} \alt 0.05 $ with Mach number ${\cal M}^R_{\infty} \ge 2.4$ the shock cone is expected to be unstable.

In this paper we explore two sizes of the black hole. We first consider a mildly small size black hole with  $r_{hor}/r_{acc}=0.1,0.26$. With these sizes the shock cone should be stable, however since they have not been shown before in 3D we present them here.  On the other hand, a second set of experiments consider a very small accretor $r_{hor}/r_{acc}=0.011,0.012,0.025$, which are in the regime of instability.

\subsection{Mildly small black hole}

In the first set of experiments we use the following parameters: $\rho_0 = 10^{-6}$, $\Gamma=4/3$, 
$c_{s\infty}$ = 0.1. The velocities are supersonic with ${\cal M}_{\infty}=3,~5$  and we use two values of the wind velocity $v_{\infty}=0.3,~0.5$, which correspond to sizes of the accretor $r_{hor}/r_{acc}=0.1$ and $0.26$ respectively. These parameters correspond to mildly small black holes with sizes bigger than those expected to show the FF instability according to \citep{foglizzo2005}. We also carried out these simulations for two values of the black hole angular momentum $a=0.5,~0.95$.

The evolution was performed using resolution $\Delta xyz=0.25M$, on a domain $[-40M,40M]^3$, where $M$ is the mass of the black hole. This is a modest resolution compared to that used in the analysis using s-lab symmetry \citep{Font1998a,donmez,cruz2012}, however this should trigger an instability faster if any. In all the cases, that is, for the two velocities, the two values of the black hole angular momentum and the five orientations of the wind we did not find the FF instability, as expected for these parameters. In Table \ref{tab:UnperturbedMach3_5} we list the set of configurations for which we track the evolution of the wind. 

\begin{figure}
\includegraphics[width=7.5cm]{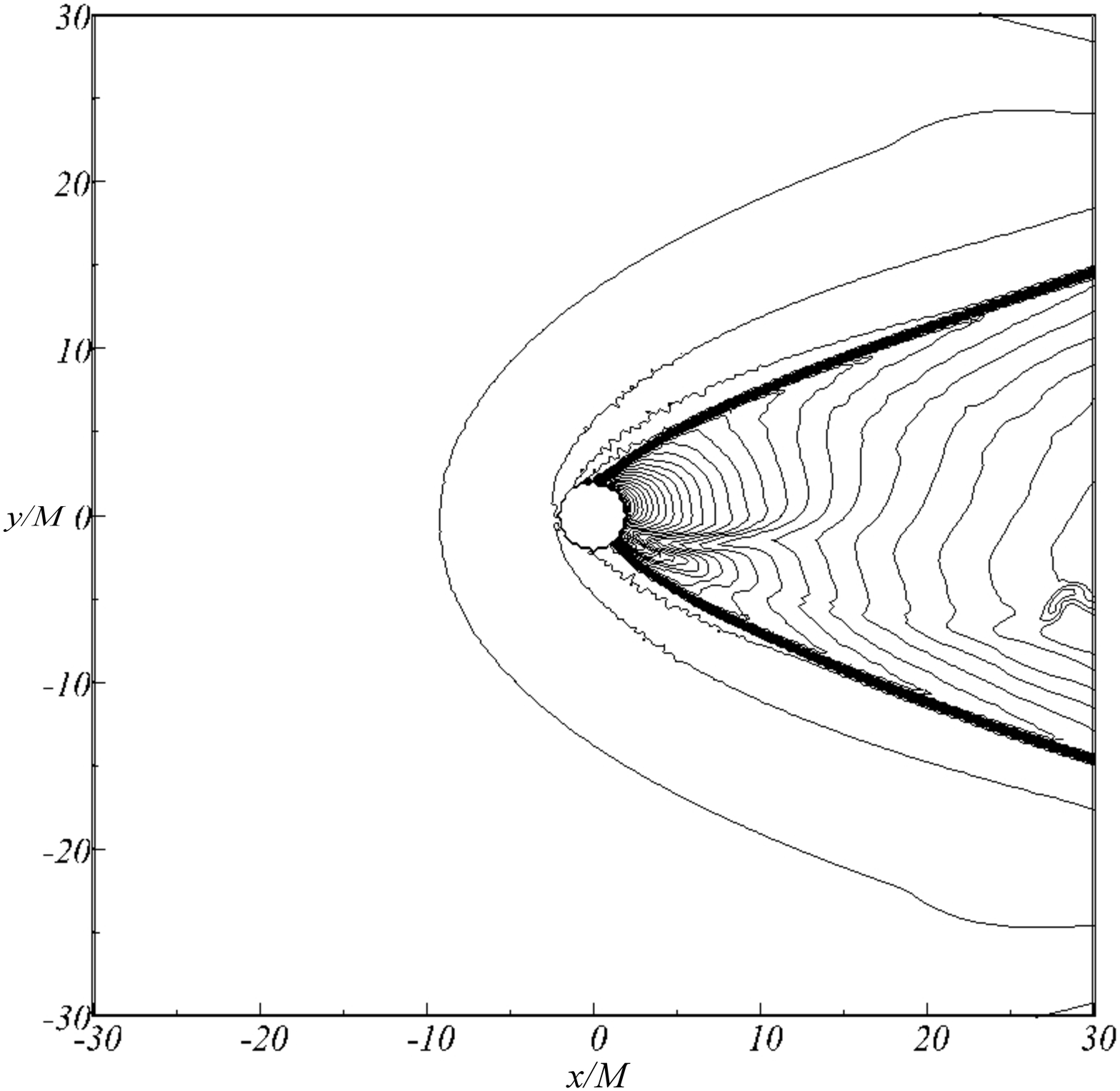}
\includegraphics[width=7.5cm]{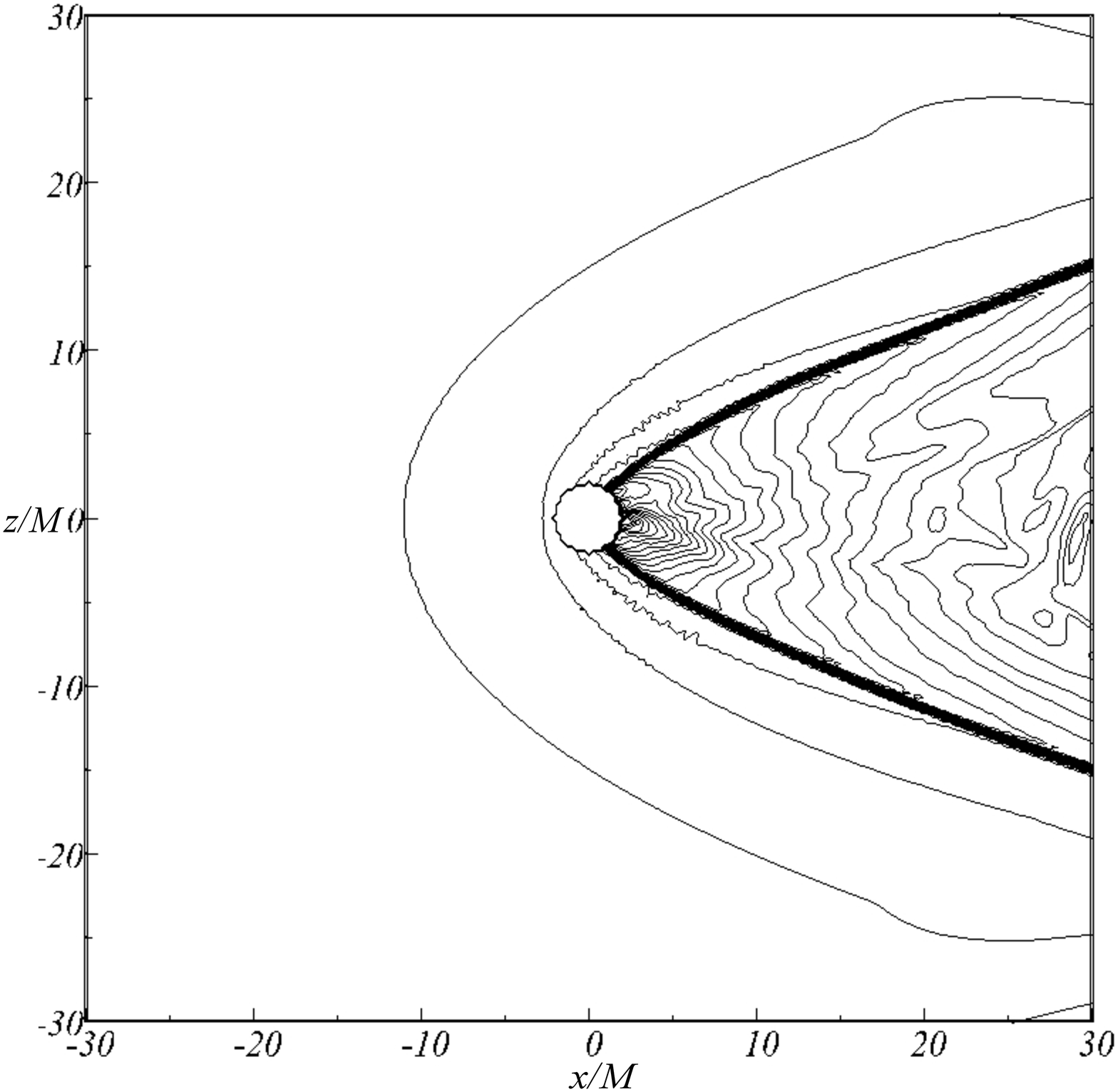}
\caption{\label{fig:FromRight} Isocurves of the density at $t=3000M$ for the mildly accretor size case $\rightarrow \Uparrow $, $a=0.95$ and $v_\infty=0.5$, $c_{s\infty}=0.1$, $r_{hor}/r_{acc}=0.26$. The thick line defining the conic shape is actually a set of nearby isocurves that show how steep the density gradient is in that region. In the top we show the cone morphology as seen at the plane $z=0$, where the dragging of the density due to the black hole rotation can be seen. In the bottom panel we show the shock cone structure on the plane $y=0$, which is the angle that is being ignored in simulations assuming s-lab symmetry.}
\end{figure}

In these simulations there are already some previously unknown results, for example, the case $ \rightarrow \Uparrow$, which is the generalization of the s-lab studies shown in for instance \citep{Font1998a,donmez,cruz2012}. In \citep{donmez} it was shown that there was a FF instability, whereas in \citep{cruz2012} the use of horizon penetrating coordinates to describe the black hole indicated that there is no such instability. However both analyses were done using s-lab symmetry and the result in 3D was still uncertain. We show for the first time, in Fig. \ref{fig:FromRight}, a snapshot of the  cone in 3D for the case with $v_\infty=0.5$, $c_{s\infty}=0.1$ and black hole angular momentum $a=0.95$. No signs of instability were found up to $t=10000M$, which is a comparable time with that in \citep{donmez,Font1998a,cruz2012}.

\begin{table}
\caption{\label{tab:UnperturbedMach3_5} Configurations analyzed for $\rho_0 = 10^{-6}$, ${\cal M}=3,~5$,  $\Gamma=4/3$, $c_{s\infty}$ = 0.1 and $v_{\infty}=0.3$. The value of $\dot{M}$ corresponds to the value measured on a spherical surface located at $r=3M$ after the process of shock cone formation has been settled down. No instability was found in all the presented cases. In order to compare how good the BHL formula is, we also show the accretion rate for each case.}
\begin{center}
\begin{tabular}{cccccc}\hline
\tableline
\tableline
Config. &  ${\cal M}=3$ & & & ${\cal M}=5$ & \\\hline
 & $\dot{M}_3 (10^{-4})$ & $\dot{M}_{3BHL}(10^{-4})$ &  & $\dot{M}_5 (10^{-4})$ &$\dot{M}_{5BHL}(10^{-4})$ \\\hline
{\bf $a=0.5$}&   \\\hline
$\uparrow \Uparrow $    &$8.29$  &  &&$3.81$\\
$\downarrow \Uparrow $  &$8.24$  &  &&$3.80$\\
$\rightarrow \Uparrow $  &$8.21$  & $4.00$ &&$3.63$ &$1.00$\\
$\swarrow \Uparrow $    &$8.27$  &  &&$3.60$\\
$\nwarrow \Uparrow $    &$8.32$  &  &&$3.61$\\\hline
{\bf $a=0.95$}&  \\\hline
$\uparrow \Uparrow $    &$8.23$ &  & &$3.64$\\
$\downarrow \Uparrow $  &$8.22$  &  &&$3.64$\\
$\rightarrow \Uparrow $  &$8.82$ & $4.00$ & &$3.63$& $1.00$\\
$\swarrow \Uparrow $    &$8.57$ &  & &$3.64$\\
$\nwarrow \Uparrow $    &$8.50$  &  &&$3.58$\\\hline
\tableline
\end{tabular}
\end{center}
\end{table}

Aside of the morphology of the shock cone that looks very stable, we monitor the stability of the accretion process by measuring the accretion rate $\dot{M}$ at a spherical surface located at $r=3M$. In all our experiments we find this rate to stabilize after an initial transient corresponding to the shock cone formation lapse. We measure this rate for the different cases and check that the slower the wind the bigger the accretion rate, as expected from the Bondi formula. We show the accretion rate also in Table \ref{tab:UnperturbedMach3_5} for the two cases ${\cal M}_{\infty}=3$ and ${\cal M}_{\infty}=5$ and for comparison we also show the values obtained using the BHL accretion rate formula \citep{edgar2004}

\begin{equation}
	\dot M_{BHL} = \frac{4\pi\rho}{(v_{\infty}^2+c_{s\infty}^2)^{3/2}};
\end{equation}

\noindent we label the results with $\dot{M}_{3NHL}$ and $\dot{M}_{5NHL}$ for the two wind velocities  considered. We point out that the accretion rate, as measured for a black hole space-time in our simulations for this size of black hole, is considerably bigger than the accretion rate calculated with the BHL formula.

\subsubsection{Particular cases}

\begin{figure}
\includegraphics[width=4cm]{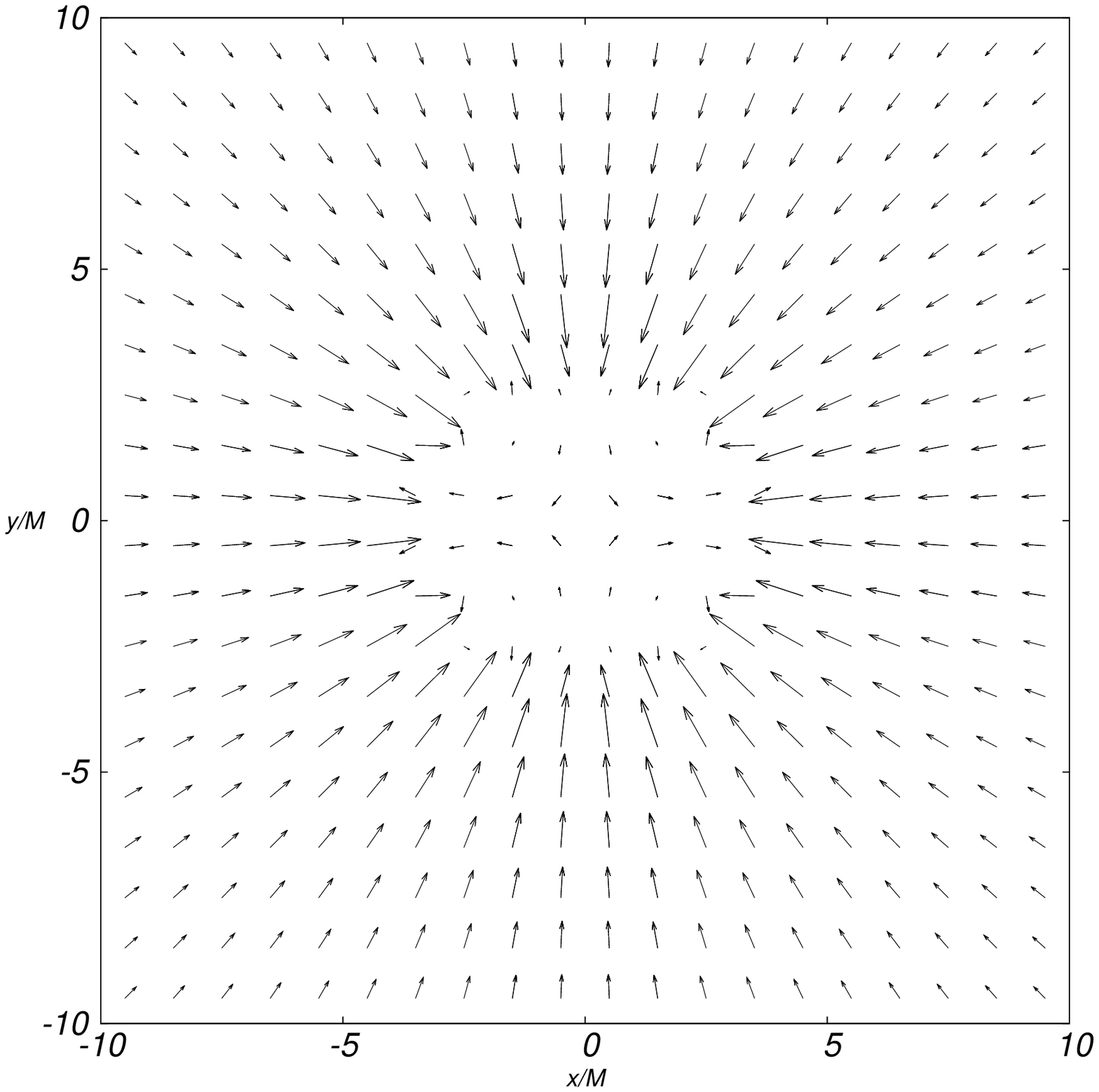}
\includegraphics[width=4cm]{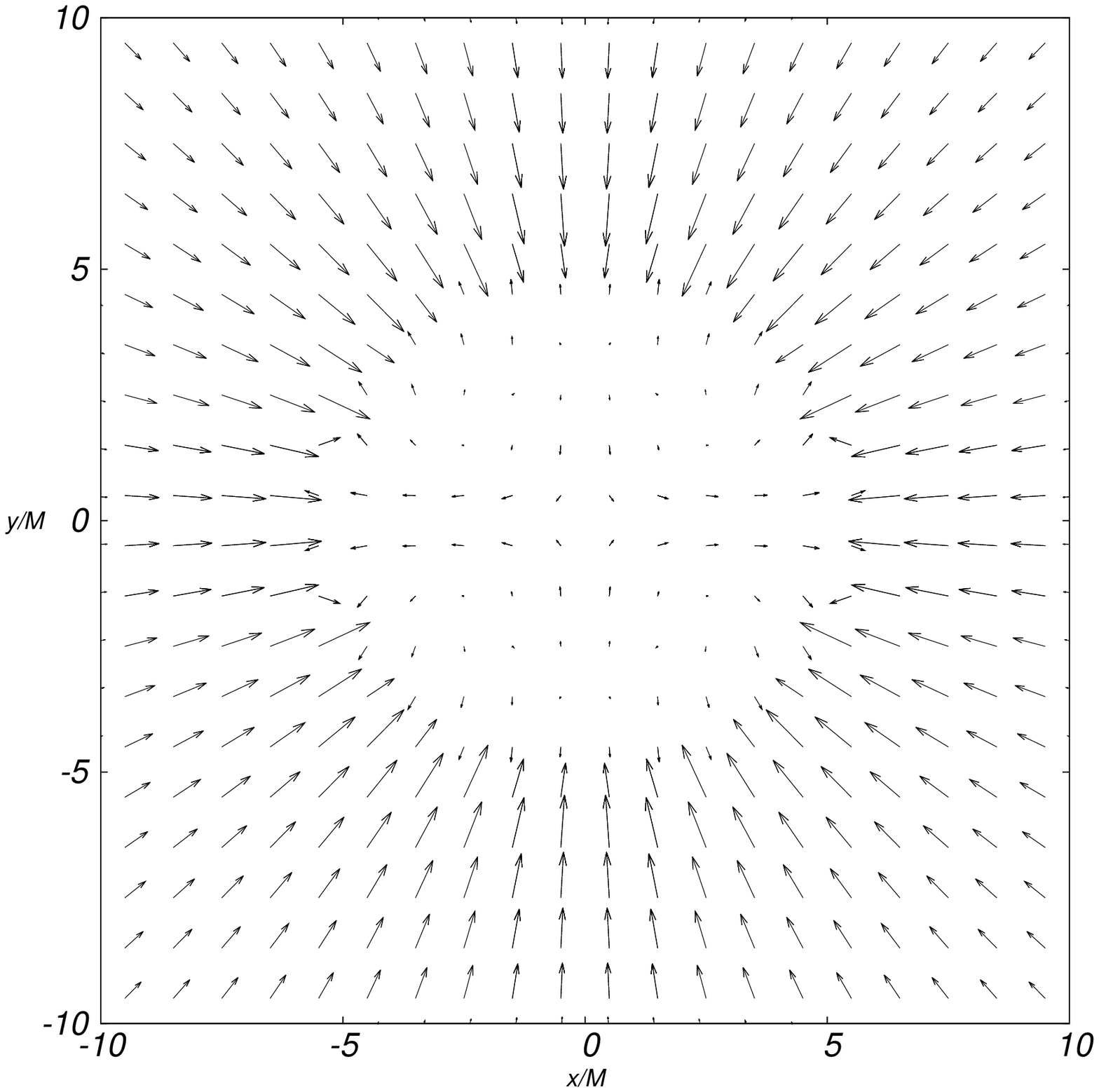}
\includegraphics[width=4cm]{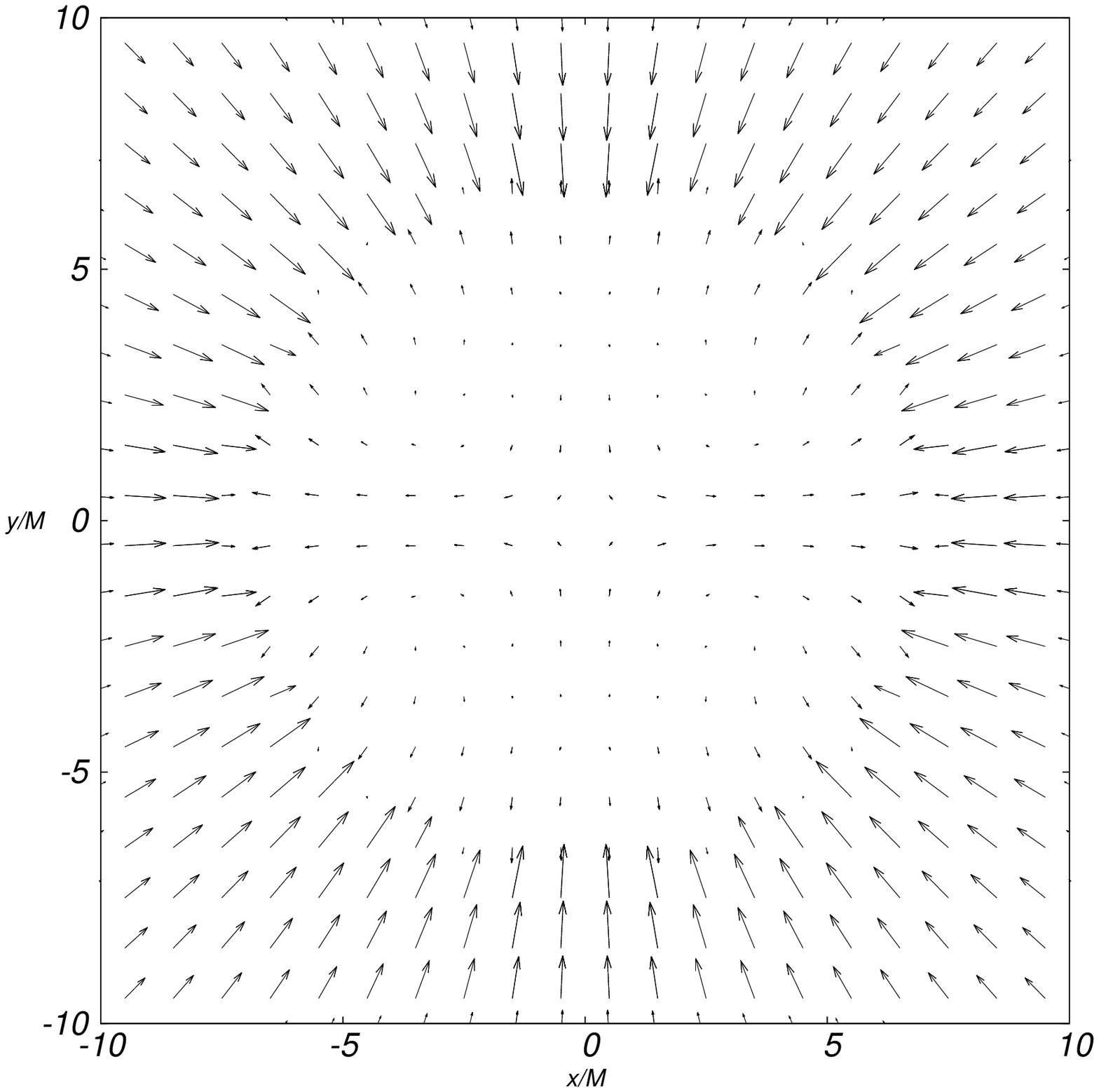}
\includegraphics[width=4cm]{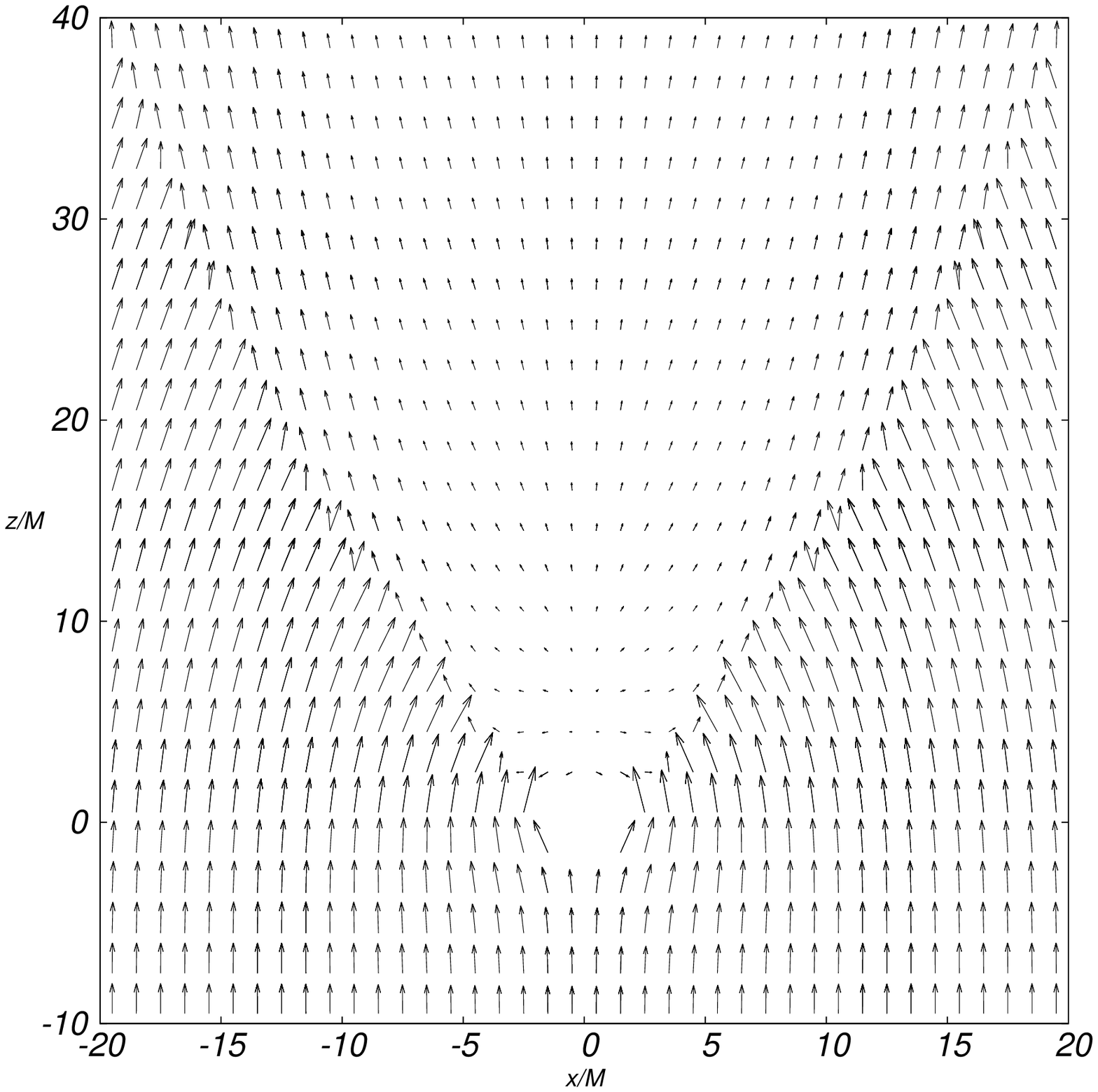}
\caption{\label{fig:vfield} Velocity field at $t=8000$M for the case $ \uparrow \Uparrow $, $v_{\infty}=0.5$, spin $a=0.95$ and $r_{hor}/r_{acc}=0.26$. The slices are as follows (top-left) is taken at $z=3$, (top-right) at $z=6$, (bottom-left) at $z=9$ and (bottom-right) at $y=0$.}
\end{figure}

Some special configurations may be interesting. For instance one may wonder whether there is a whirlpool behind the black hole when the wind moves parallel or antiparallel to the axis of rotation. In Fig. \ref{fig:vfield} we show slices of the velocity field at three planes of $z=$constant,  and a lateral view of the process at $t=8000M$. In this case the black hole size is $r_{hor}/r_{acc}=0.26$ and no dragging effects are noticeable in the velocity field due to the rotation of the black hole. 
What can actually be seen is the convergence of the stream lines at the shock cone boundary.

\begin{figure}
\includegraphics[width=7.5cm,height=7.5cm]{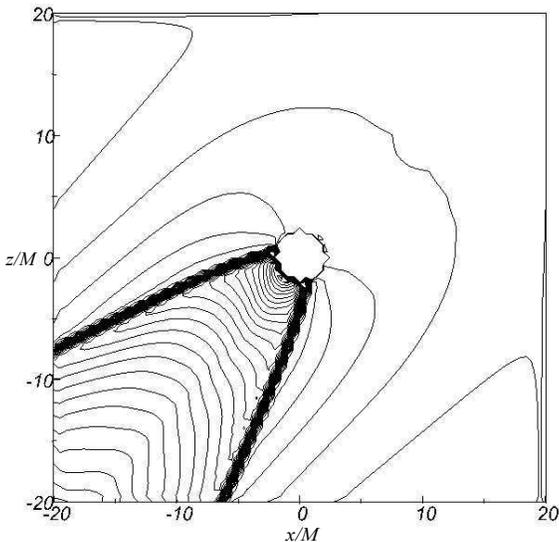}
\caption{\label{fig:diagonal} We show a snapshot of the density isocurves for the diagonal case $\swarrow \Uparrow $, $a=0.5$ and $v_\infty=0.5$, $c_{s\infty}=0.1$, $\Gamma=4/3$ and $r_{hor}/r_{acc}=0.26$.}
\end{figure}

Another case that has never been shown in 3D is the accretion of a diagonal wind. We show a snapshot of the shock cone density at $t=5000M$ in Fig. \ref{fig:diagonal}, for the wind orientation $\swarrow \Uparrow$.

\subsection{Very small black hole}

A second set of tests consists in the evolution of the wind for a smaller black hole, in particular we set three cases: $r_{hor}/r_{acc}=0.011,0.012,0.025$. Being this number smaller that the threshold 0.05, the FF instability is expected to happen as suggested in \citep{foglizzo2005}. According to the analysis in \citep{foglizzo2005}, even axisymmetric simulations should become unstable when considering $\Gamma = 5/3, 4/3$ and  ${\cal M}_{\infty} = 2.4,~3.0$. We evolved these particular cases in 3D, for all the wind orientations described before, and presented in Table \ref{tab:slowUnperturbedMach3_5}. In this case we only perform the simulations for the bigger black hole angular momentum $a=0.95$. We use the same resolution $\Delta xyz = 0.25M$, however, because a domain containing at least a sphere of radius $r_{acc}$ is required, we used the domain $[-100M,100M]^3$. Once again we did not find the FF instability after $10000 M$. We point out that unlike the mildly small accretor size, the accretion rate for this size of accretor is considerably smaller than the accretion calculated with the BHL formula.

\begin{figure}
\includegraphics[width=7.5cm]{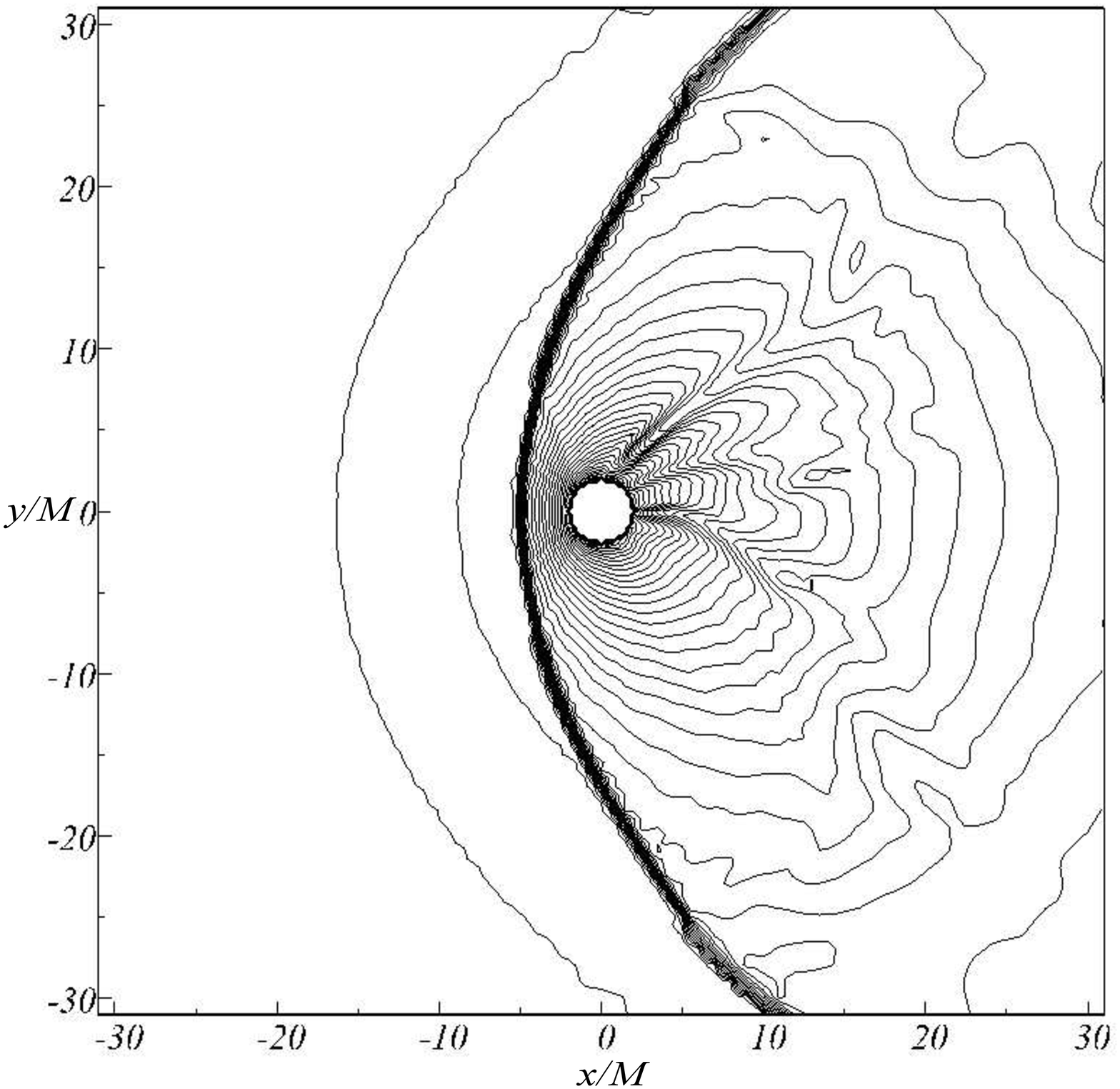}
\includegraphics[width=7.5cm]{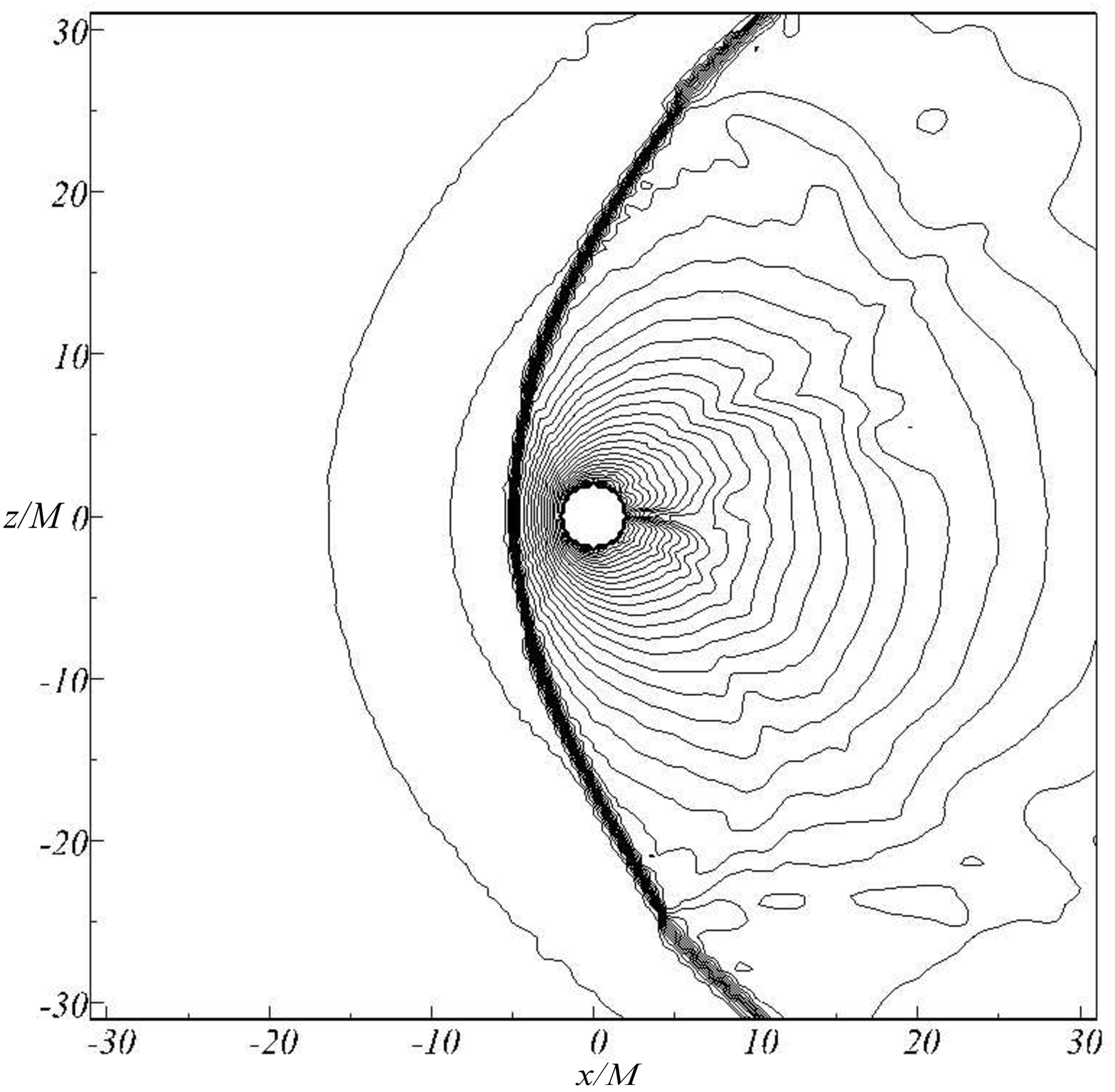}
\caption{\label{fig:gamma5/3} Isocurves of the density at time $t=8000M$, for the case $\leftarrow \Uparrow $, $a=0.95$ and $v_\infty=0.1$, $c_{s\infty}=0.04$, $\Gamma=5/3$, corresponding to a small black hole with $r_{hor}/r_{acc}=0.011$. We show two views, at the $z=0$ and $y=0$ planes. This case is expected to be unstable because the shock is detached from the black hole, and the accretor size is small enough. Even though, we did not encounter signs of FF instability until $t=10000M$.}
\end{figure}

\begin{figure}
\includegraphics[width=7.5cm,height=7.5cm]{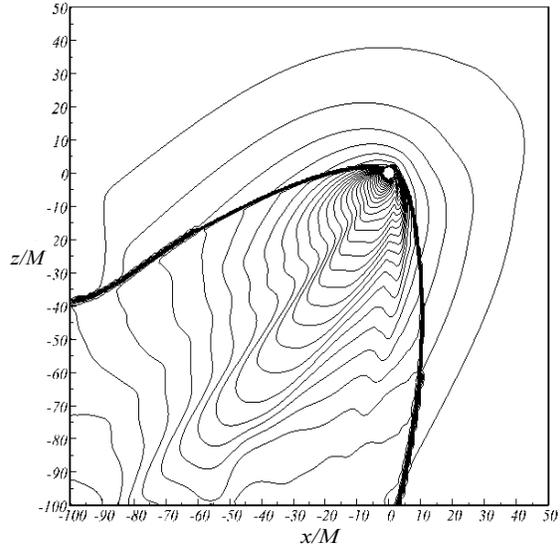}
\caption{\label{fig:diagonal2} Density isocurves for the diagonal case $\swarrow \Uparrow $, $a=0.95$, $\Gamma=4/3$ and $r_{hor}/r_{acc}=0.012$. The snapshot is taken at $t=3800M$.}
\end{figure}

In order to illustrate how the dragging effects due to the rotation of a small black hole on the fluid are more significant compared to the bigger black hole case, we show in Fig. \ref{fig:gamma5/3} a snapshot of the density for the case with orientation $ \rightarrow \Uparrow $, $r_{hor}/r_{acc}=0.011$, $\Gamma=5/3$ and ${\cal M}=2.4$ from Table \ref{tab:slowUnperturbedMach3_5}. In Fig. \ref{fig:diagonal2} we also show a snapshot on a vertical plane of the diagonal case, which shows a more dragged fluid than in the case of Fig. \ref{fig:diagonal}.


\begin{table}
\caption{\label{tab:slowUnperturbedMach3_5} Configurations analyzed for $\rho_0 = 10^{-6}$ and $a=0.95$. The value of $\dot{M}$ corresponds to the value measured on a spherical surface located at $r=3M$ after the process of shock cone formation has been settled down. None of the cases shows instability. In order to compare how good the BHL formula is, we also show the accretion rate for each case. Unlike in the case of mildly small black holes, in this case the numerical accretion rate is about the same independently of the orientation of the wind.}
\begin{center}
\begin{tabular}{cccc}\hline
\tableline
\tableline	
$\Gamma=5/3$ & &$r_{hor}/r_{acc}=0.011$ &${\cal M}=2.4$\\\hline
&&$\dot{M}_{2.4} (\times 10^{-2})$ & $\dot{M}_{BHL}(10^{-2})$\\\hline\hline
$\uparrow \Uparrow $     &&\\
$\downarrow \Uparrow $    &&\\
$\rightarrow \Uparrow $     &&$0.14$&$1.00$\\
$\swarrow \Uparrow $       &&\\
$\nwarrow \Uparrow $       &&\\\hline\hline
$\Gamma=4/3$ & &$r_{hor}/r_{acc}=0.012$ &${\cal M}=2.5$ \\\hline
&&$\dot{M}_{2.5} (\times 10^{-2})$ & $\dot{M}_{BHL}(10^{-2})$\\\hline\hline
$\uparrow \Uparrow $       &&\\
$\downarrow \Uparrow $     & &\\
$\rightarrow \Uparrow $     && $0.48$ &$1.00$\\
$\swarrow \Uparrow $       &  &\\
$\nwarrow \Uparrow $       &  &\\\hline\hline
$\Gamma=4/3$ & &$r_{hor}/r_{acc}=0.025$ &${\cal M}=3$ \\\hline
&& $\dot{M}_{3.0} (\times 10^{-2})$ & $\dot{M}_{BHL}(10^{-3})$\\\hline\hline
$\uparrow \Uparrow $       &&\\
$\downarrow \Uparrow $     &&\\
$\rightarrow \Uparrow $     &&$0.34$&$3.72$\\
$\swarrow \Uparrow $       &&\\
$\nwarrow \Uparrow $       &&\\\hline
\tableline
\end{tabular}
\end{center}
\end{table}

\subsection{Perturbed Bondi-Hoyle accretion}

As we did not find any unstable case, in addition to the analysis of simply tracking the evolution of the gas and expect the numerical error to trigger the FF instability, we designed a numerical experiment in which we evolve the gas until the shock cone forms, and then apply a perturbation to the velocity field of the form

\begin{equation}
 v^i = v^i + k\delta(t)
\end{equation}

\noindent where $k \ge 0$ is a random number between $0$ and $A$. We applied this random perturbation perpendicular to the direction of the shock cone and expected for it to trigger the instability.  We did this experiment using the two extreme relative accretor sizes in this paper $r_{hor}/r_{acc}=0.011$ and  $r_{hor}/r_{acc}=0.26$, and all the orientations we have considered here. In all the cases no instability was triggered for all the directions of the wind. 

\begin{figure}
\includegraphics[width=7.5cm]{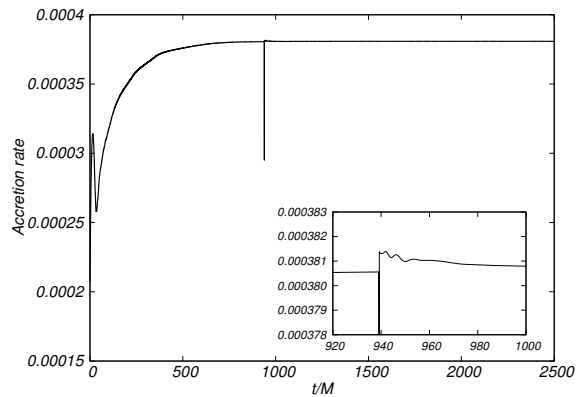}
\caption{\label{fig:accrateperturbed} Accretion rate for the perturbed case $\rightarrow\Uparrow $, $a=0.95$ and $v_\infty=0.5$, $c_{s\infty}=0.1$, $\Gamma=4/3$ and perturbation amplitude $A=0.2$. In the inset we show the instant when the perturbation is applied by time $t=939M$.}
\end{figure}

\begin{figure}
\includegraphics[width=4cm,height=4cm]{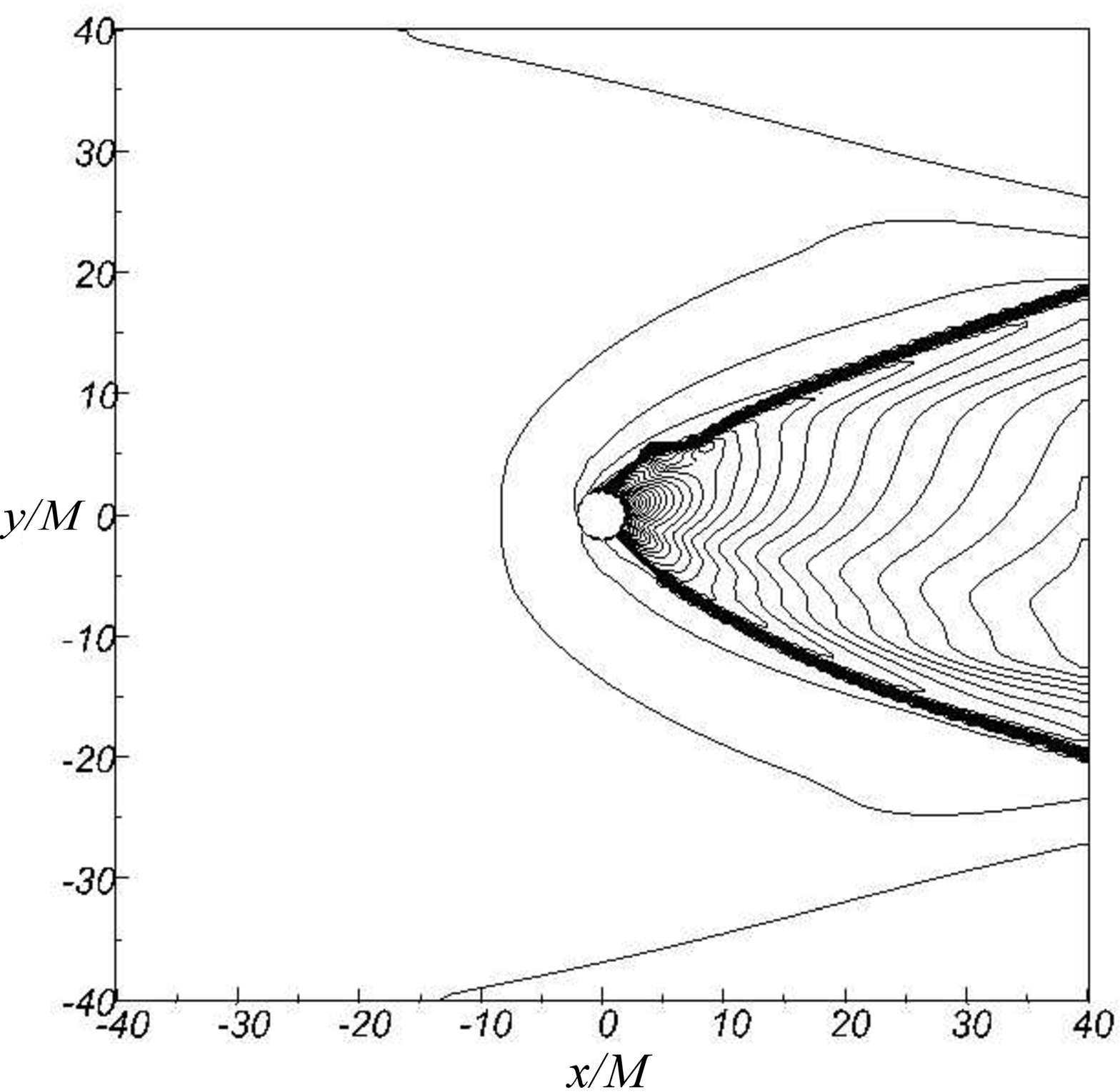}
\includegraphics[width=4cm,height=4cm]{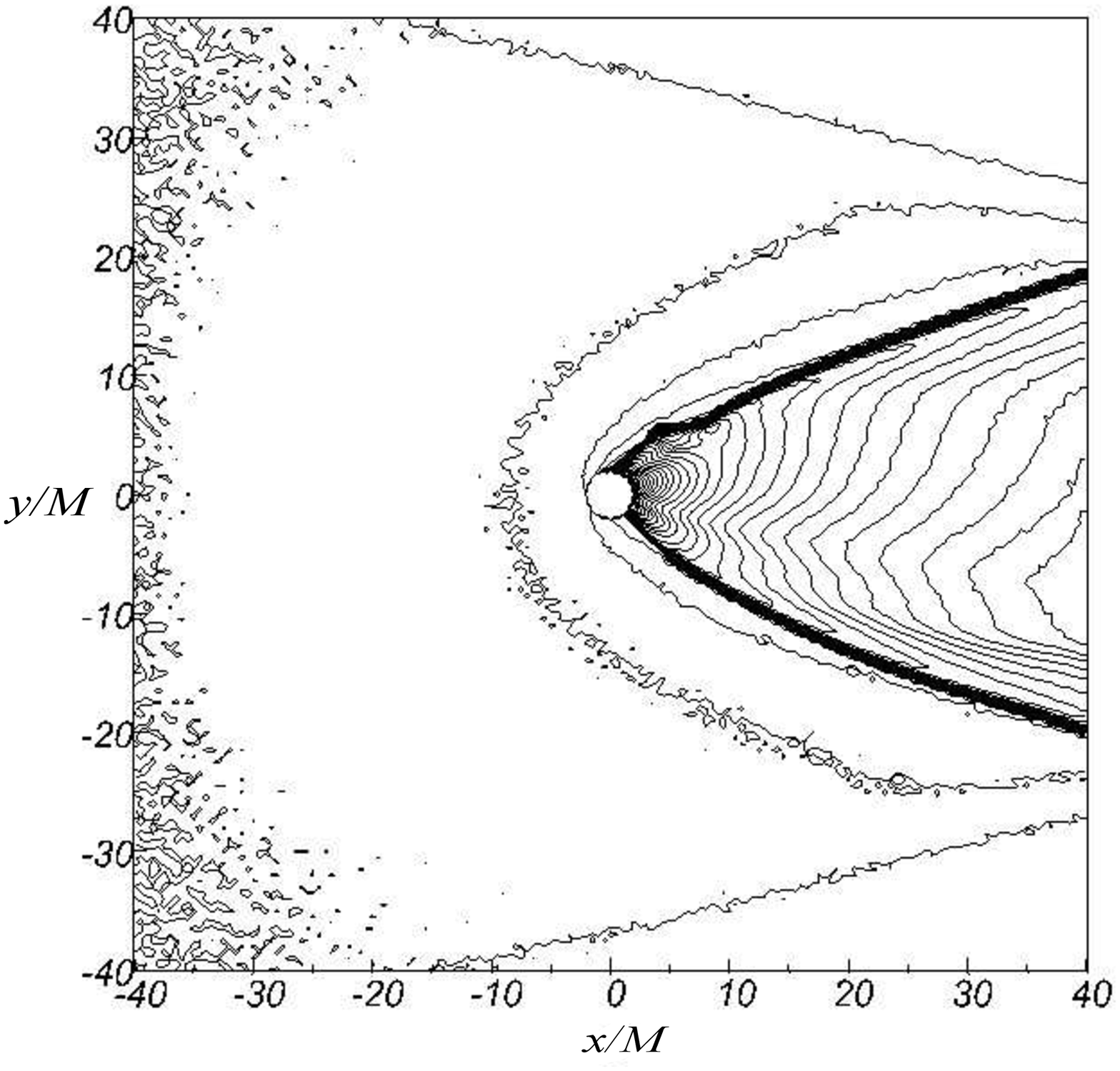}
\includegraphics[width=4cm,height=4cm]{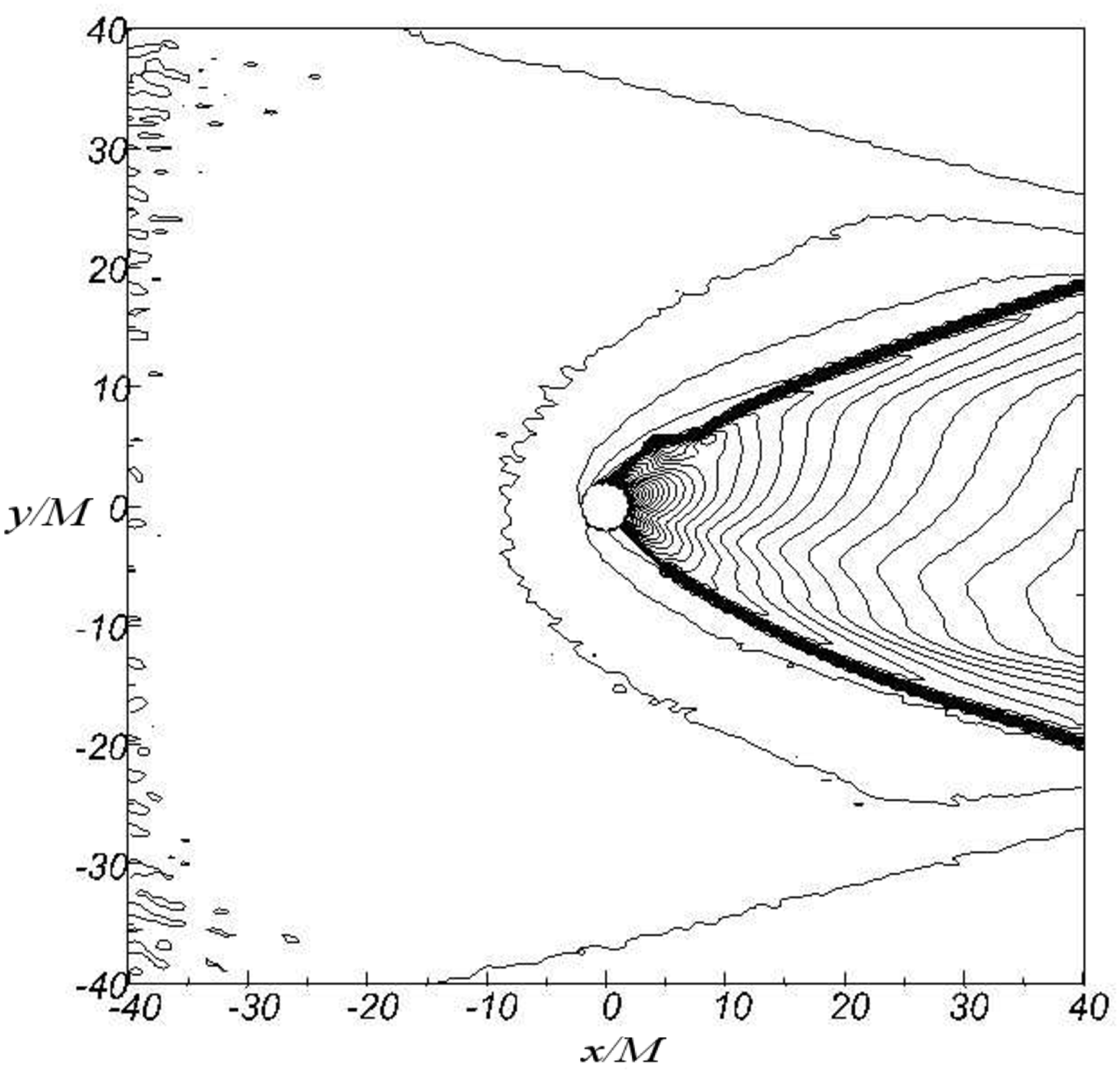}
\includegraphics[width=4cm,height=4cm]{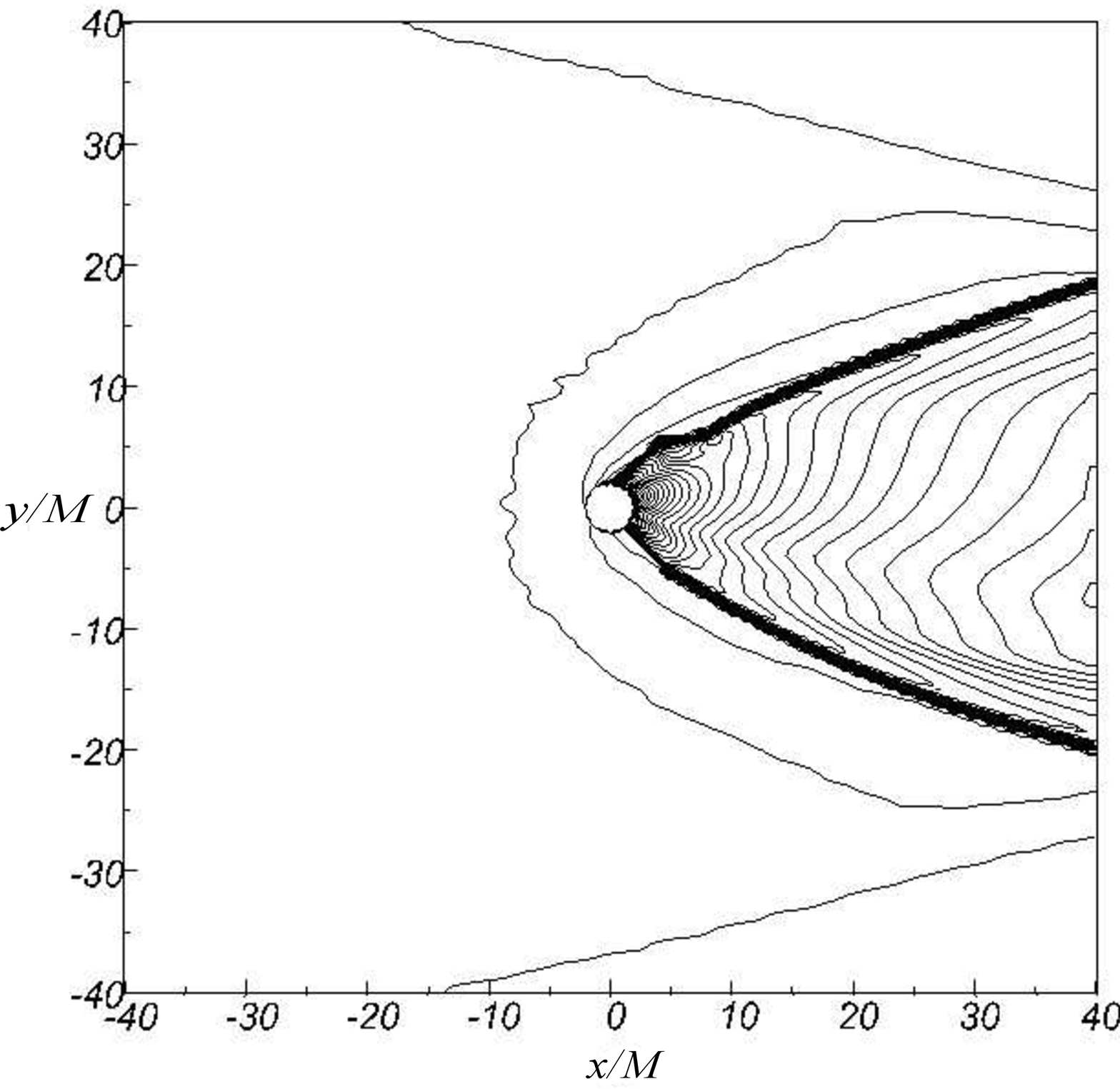}
\caption{\label{fig:vfieldpert} We show snapshots of the density for a perturbed case with orientation $ \rightarrow\Uparrow $, $a=0.95$ and $v_\infty=0.5$, $c_{s\infty}=0.1$, $\Gamma=4/3$ with a perturbation amplitude $A=0.2$. In the top-left panel we show the density before applying the perturbation. In the top-right we show the morphology at $t=939M$ when the perturbation was applied. In the bottom panels we show how the sock cone restores at times $t=945M$ and $t=955M$.}
\end{figure}

\begin{figure}
\includegraphics[width=4cm,height=4cm]{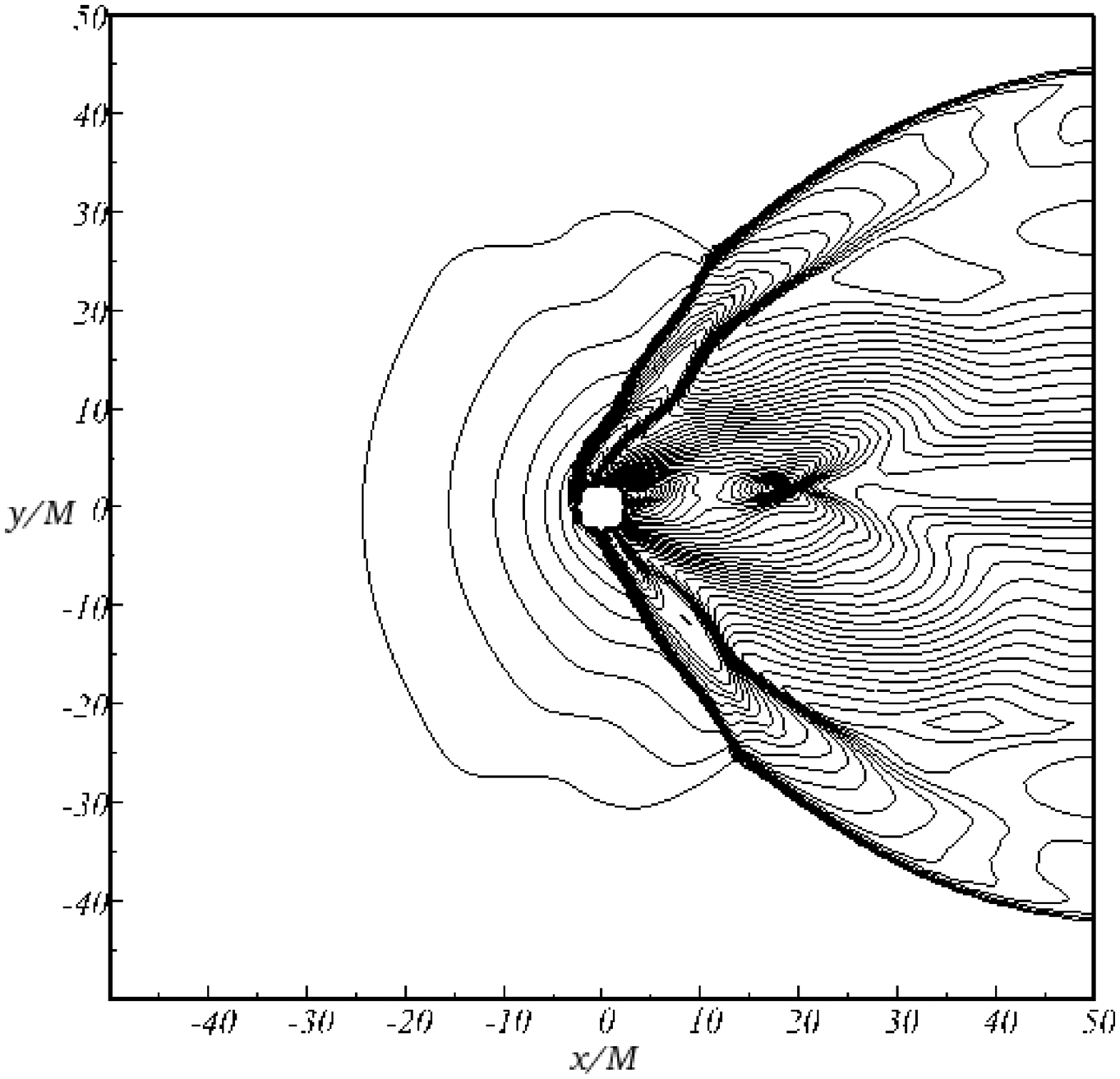}
\includegraphics[width=4cm,height=4cm]{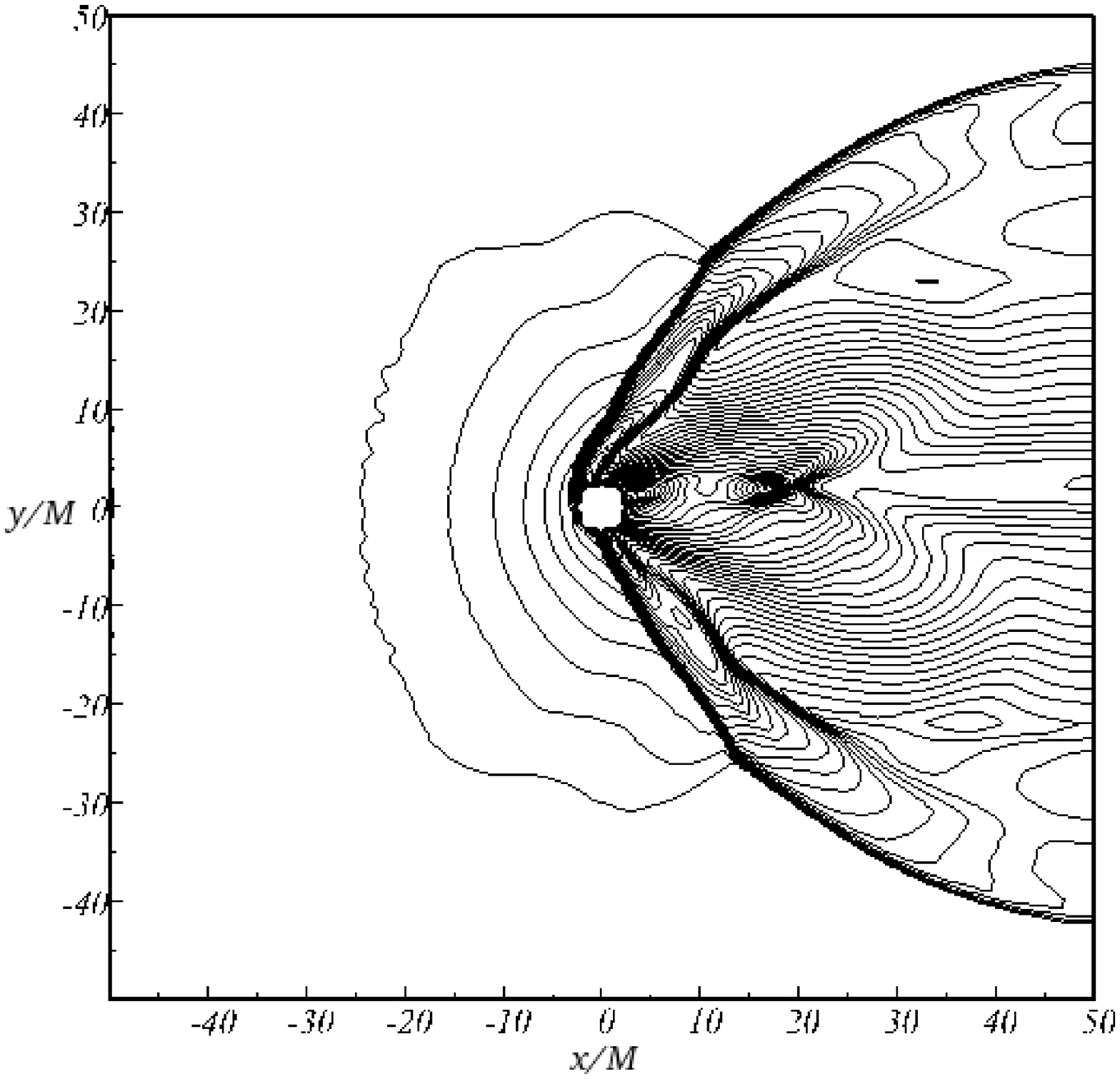}
\includegraphics[width=4cm,height=4cm]{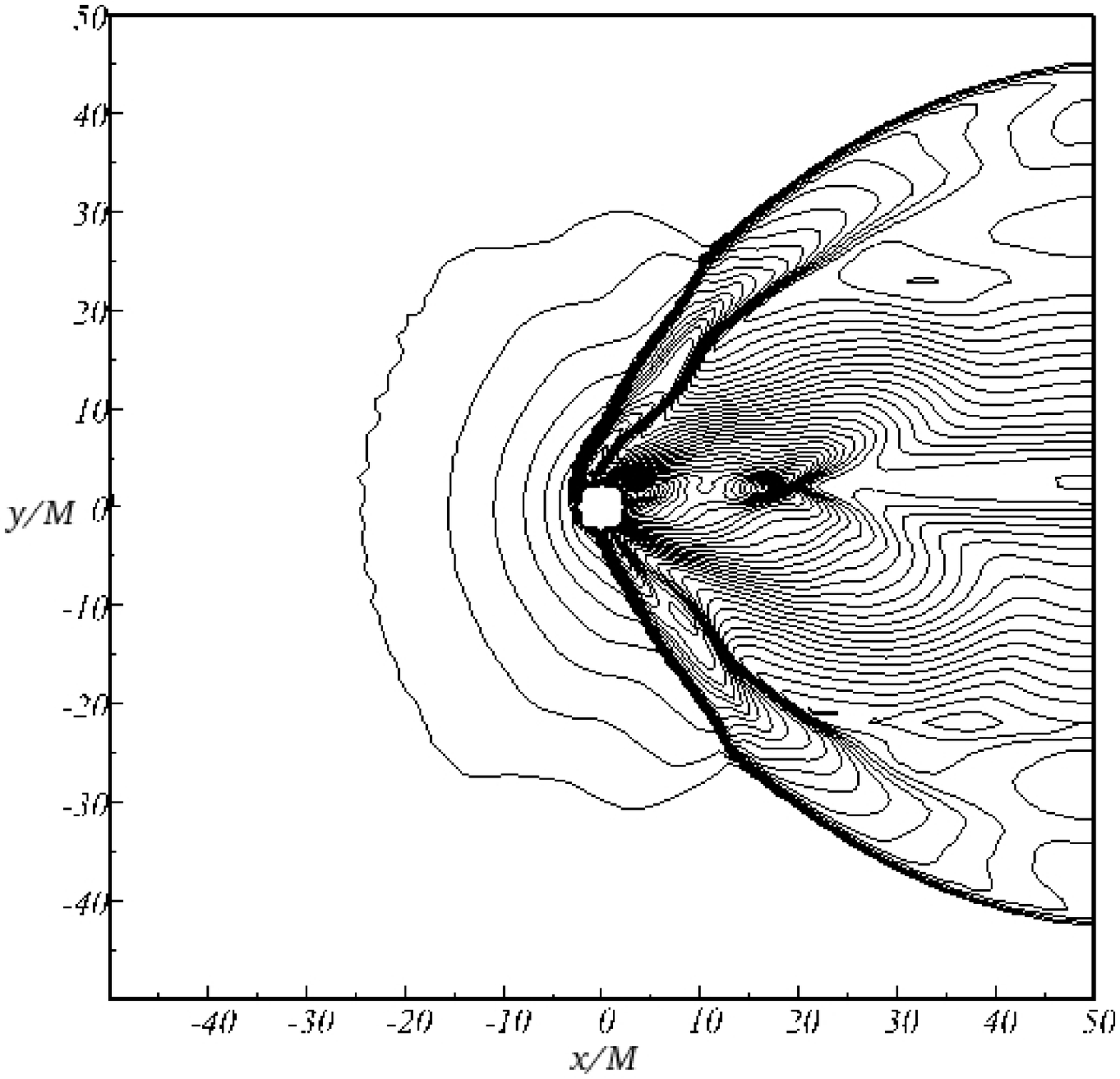}
\includegraphics[width=4cm,height=4cm]{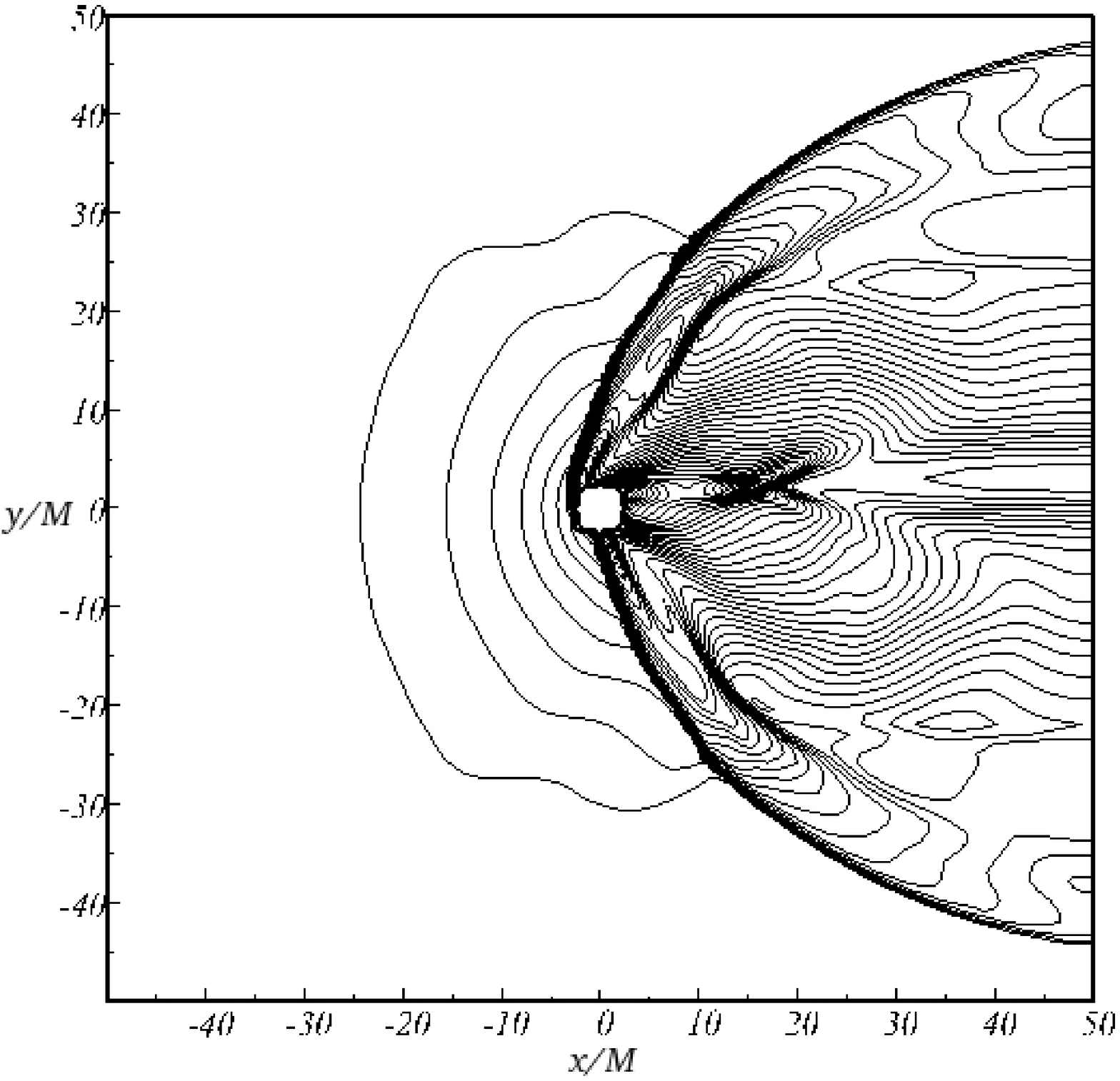}
\caption{\label{fig:vfieldpert2} We show snapshots of the density for a perturbed case with orientation $ \rightarrow\Uparrow $, $a=0.95$, $r_{hor}/r_{acc}=0.011$, $v_\infty=0.1$, $c_{s\infty}=0.04$, $\Gamma=4/3$ with a perturbation amplitude $A=0.2$. In the top-left panel we show the density before applying the perturbation. In the top-right we show the morphology at $t=1500M$ when the perturbation was applied. In the bottom panels we show how the sock cone restores at times $t=1530M$ and $t=1600M$.}
\end{figure}

In Fig. \ref{fig:accrateperturbed} we show the accretion rate for the perturbed case, where a feature can be seen at the moment at which we apply the perturbation. We used $A=0.2$, which means that the velocity perpendicular to the shock cone has been added randomly even one fifth of the speed of light and nevertheless the flux restores and becomes stable. In Figs. \ref{fig:vfieldpert} and \ref{fig:vfieldpert2} we show snapshots of the density before, during and after the perturbation is applied, and show how the morphology of the cone restores.

\subsection{Influence of numerical methods}

We wanted to keep the physical system as free as possible of numerical sophistication in order to avoid undesirable effects. For this, in the production runs we used a single refinement level, because we wanted to avoid the numerical dissipation required to damp instabilities at refinement level interfaces, which could hide the potential unstable modes of the shock cone. This has been discussed as a potential damping of the instability in FF type of phenomena \citep{foglizzo2005}.

\begin{figure}
\includegraphics[width=7.5cm]{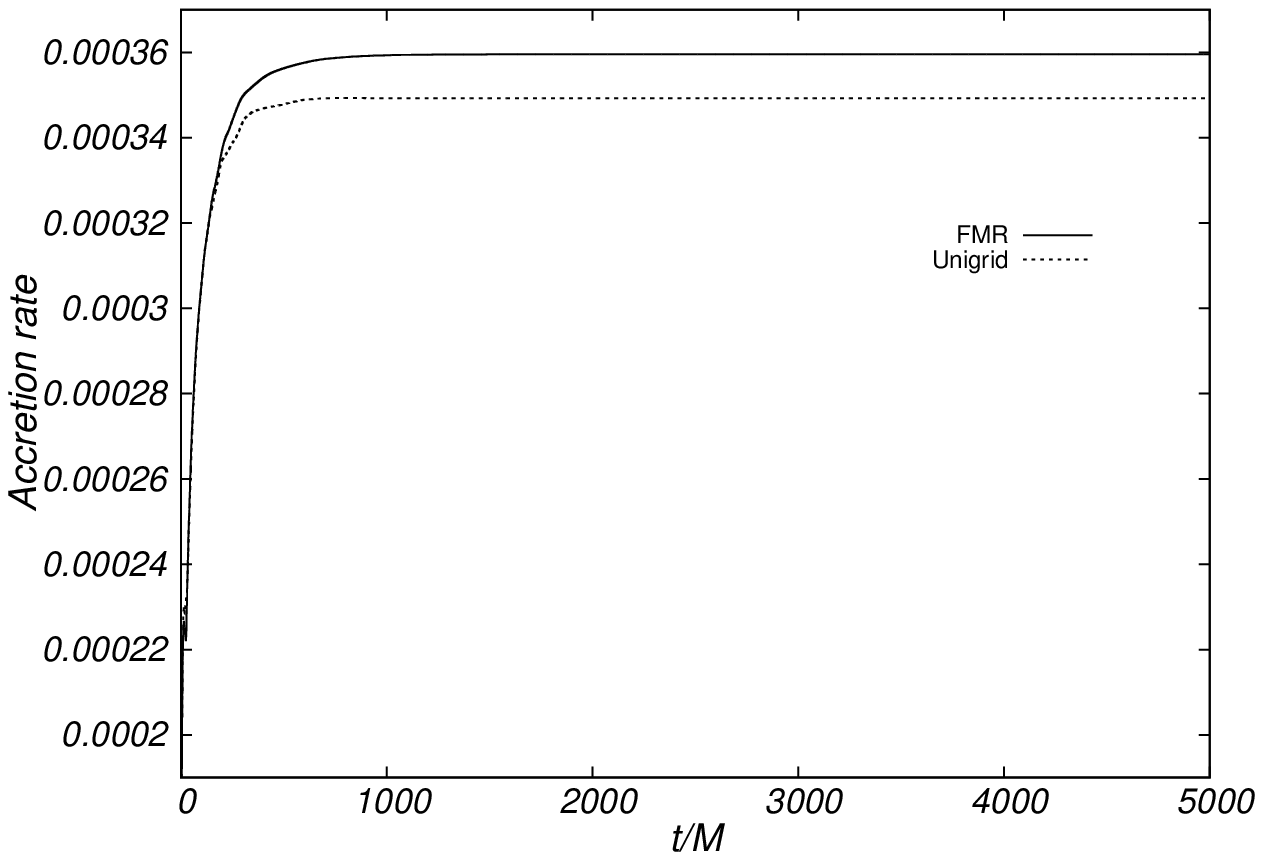}
\includegraphics[width=7.5cm]{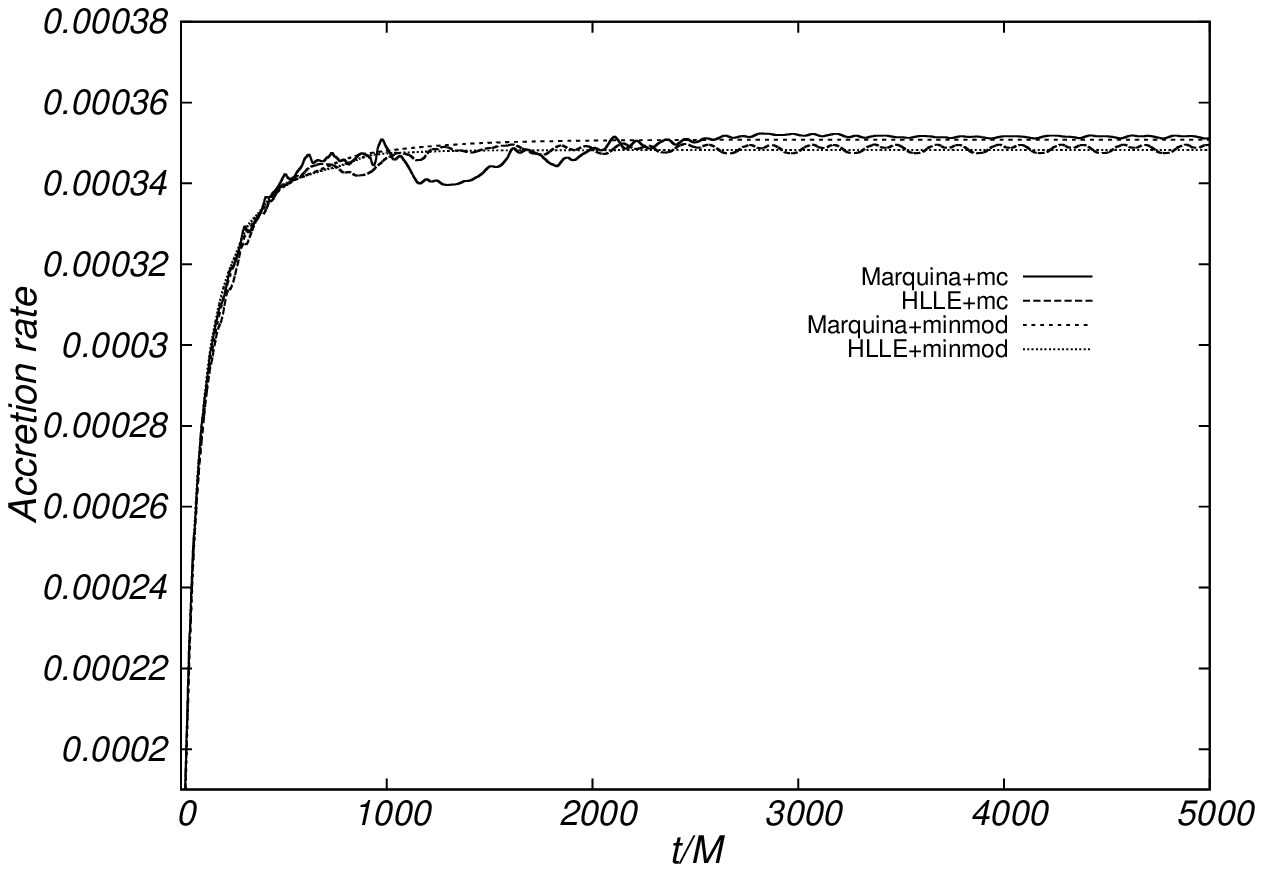}
\caption{\label{fig:amrvsuni} (Top) Accretion rate comparison for the case $ \rightarrow \Uparrow $, $a=0.95$, ${\cal M}=5$, between two runs, one using two refinement levels, and the other one using unigrid mode. In the unigrid mode, the whole domain $[-20M,20M]^3$ is covered using resolution $0.25M$ in all directions. In the case of two refinement levels, the full domain is covered with a resolution $0.5M$ and a small box with resolution $0.25M$ is set in the domain $[-10M,10M]^3$. The accretion rate is measured at a sphere of radius $3M$ and the differences are of $3\%$, which eventually might be important. (Bottom) Accretion rate comparison for the case $ \rightarrow \Uparrow $, $a=0.95$, ${\cal M}=5$, between four runs, using Marquina and HLLE as Riemann solvers, both with MC and minmod as reconstruction methods, using the domain $[-20M,20M]^3$ in unigrid mode covered with resolution $0.25M$. The difference in the accretion rate among the four combination of numerical methods lies within 1 \%.}
\end{figure}

In order to show the effects due to the use of FMR we show in Fig. \ref{fig:amrvsuni} the accretion rate when using FMR with two refinement levels and when using a single refinement level. The deviation is of the order of 3\% in the mass accretion rate. Thus, even though FMR allows to execute long runs more efficiently the accuracy is not a good as with unigrid, which is important in long runs.

We also verified our results on the stability of the shock cone in all our runs using various combinations of approximate Riemann solvers and reconstruction methods. We evolved all the configurations in the Tables using four combinations, namely: 1) Marquina with MC and minmod and 2) HLLE with MC and minmod. In all the cases we found the shock cone to be stable, in all the described orientations and velocities of the wind. In Fig. \ref{fig:amrvsuni} we compare the accretion rate estimated using the four combinations of fluxes and reconstructors for a particular case. The difference among the accretion rate lies within 1 \%, which is smaller than the error introduced when using FMR.

\section{Discussion and conclusions}
\label{sec:conclusions}

We presented the supersonic accretion of a wind onto black holes in 3D and studied whether or not there is a FF instability using a range of parameters expected to show this instability. We track the formation of the shock cone considering a high black hole spin and a general set of orientations of the wind with respect to the orientation of the BH axis of rotation.

We focus on two regimes, one in which the relative accretor size is $r_{hor}/r_{acc} = 0.1,~0.26$ where the FF instability is not expected, however we present the accretion in three dimensions. A second regime in which the relative accretor size is $r_{hor}/r_{acc} = 0.011,~0.012,~0.025$, which should show the FF instability according to  \citep{foglizzo2005}. In both cases we did not see the instability during the evolution time where it was expected to happen.

A first set of experiments assumes the numerical errors could trigger the instability of the shock cone. Since we did not find the instability we ran a second set of experiments in which we applied a transverse perturbation to the shock cone and also failed to see the instability.

Our production runs were done in unigrid mode in order to avoid the numerical artifacts induced by mesh refinement, in particular the artificial dissipation. We illustrate with a particular case, the estimated deviation from unigrid mode induced by the use of FMR.

Another particular ingredient of our simulations is that the description of the black hole space-time uses Kerr-Schild horizon penetrating coordinates, and we use excision inside the event horizon. This inhibits  the propagation of numerical errors generated at the inner boundary. As described in \citep{cruz2012} this simple fact makes the difference between stability and instability of the shock cone when the accretor is a black hole. Unfortunately this technique can only be applied to black holes and not to other compact objects.

One more result is that the BHL formula is not accurate for both of our regimes. For the cases where $r_{hor}/r_{acc}=0.1,~0.26$, the accreted material on a surface located at $3M$ is bigger than twice the accretion predicted by the BHL formula, whereas for the cases of small black holes with $r_{hor}/r_{acc}$ ranging from 0.011 to 0.025, the accretion rate ranges from nearly $\sim 0.15$ to $\sim 0.5$ times those calculated with the BHL formula, depending on the value of $\Gamma$ and how supersonic the wind is.

Given the rich combination of wind orientations and values of the black hole angular momentum explored, we can safely conclude that even for our most extreme case, with a wind moving at ${\cal M}_{\infty} = 2.4$ and black holes as small as $r_{hor}/r_{acc} =0.011$, the shock cone is stable. And for faster winds with ${\cal M}_{\infty} = 3$ and $r_{hor}/r_{acc} =0.025$ within the regime of expected instability there are no signs of instability either.

\appendix

The implementation of the relativistic hydrodynamics, both in special and general relativity of the ETK has been tested severely \citep{ETK,ETK2012}. However we have added a module that deals with specific boundary conditions that needs to be tested. We present a test relevant to the results presented here, the exact solution of radial accretion of an ideal gas onto a Schwarzschild black hole known as Michel solution \citep{papa}.  The first step to construct the exact solution is to integrate the evolution equations in spherical symmetry $\partial_\mu(\sqrt{-g}\rho u^\mu) = 0$ and $\nabla_\mu T^{\mu}_{\;0} = 0$,

 
\begin{equation}
\sqrt{-g}\rho u^r = c_1 , ~~~
\sqrt{-g}\rho h u^r u_0 = c_2.
 \label{eq:constante1}
\end{equation}
 
\noindent The result from dividing both equations and differentiate with respect to $\rho$, $u^r$ and $r$ is 
 
 
 \begin{equation}
 \frac{1}{\rho}d\rho+ \frac{1}{u^r}du^r + \frac{1}{\sqrt{-g}}\partial_r\sqrt{-g}dr  = 0 ,~~~ 
 \frac{2\partial_\rho h}{h}d\rho + \frac{2u^r}{u_0^2}du^r - \frac{\partial_r g_{00}}{u_0^2}dr  = 0, \label{eq:urho2}
 \end{equation}

 \noindent after we substitute $d\rho/\rho$ in equations (\ref{eq:urho2}) and reorder therms we obtain $\frac{du^r}{u^r}\left[V^2  - \left(\frac{u^r}{u_0}\right)^2\right] + \frac{dr}{r}\left[2V^2 - \frac{M}{ru_0^2}\right]=0$, where $V^2 \equiv \frac{d \ln(\rho h)}{d\ln \rho } - 1$. In order to have solution for $u^r$ and $r$, the condition must satisfy at the same time the relations



 \begin{equation}
 V^2 - \left(\frac{u^r}{u_0}\right)^2  = 0, ~~~
 2V^2 - \frac{M}{ru_0^2}  = 0, \label{eq:critic1}
 \end{equation}

\noindent the result of solving  (\ref{eq:critic1}) are the critical points $u_c^r= \frac{M}{2r_c}$ and $V^2_c = \frac{u_c^r}{(u^r_c)^2 - (g_{00})_c }$.
 

On the other hand if we consider the gas obeys a polytropic equation of state $P=\kappa\rho^\Gamma$, and if a value of $\Gamma$ and $\rho_c$ is fixed,  we can calculate the polytropic constant $\kappa$.
 
Once we know the critical values $u^{r}_{c}$, $V_c$  and $\kappa$, the constants $c_1$ and $c_2$ can be calculated and equations (\ref{eq:constante1}) can be solved. Although we have constructed an exact solution, the system can not be solved analytically and for this reason we programed a Newton-Raphson root finder to perform this task.

\begin{figure}
	\begin{center}
\includegraphics[width=6.5cm,height=6.5cm]{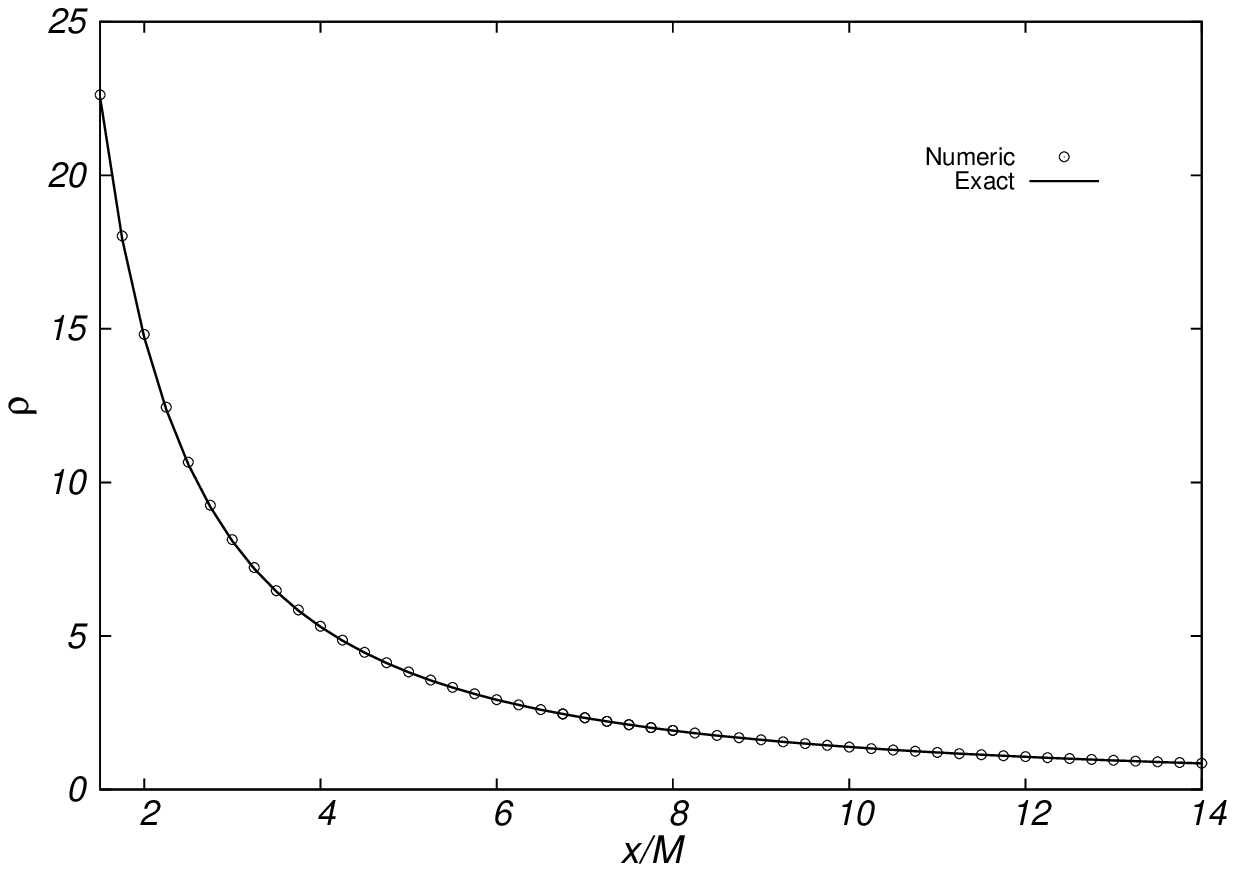}
\includegraphics[width=6.5cm,height=6.5cm]{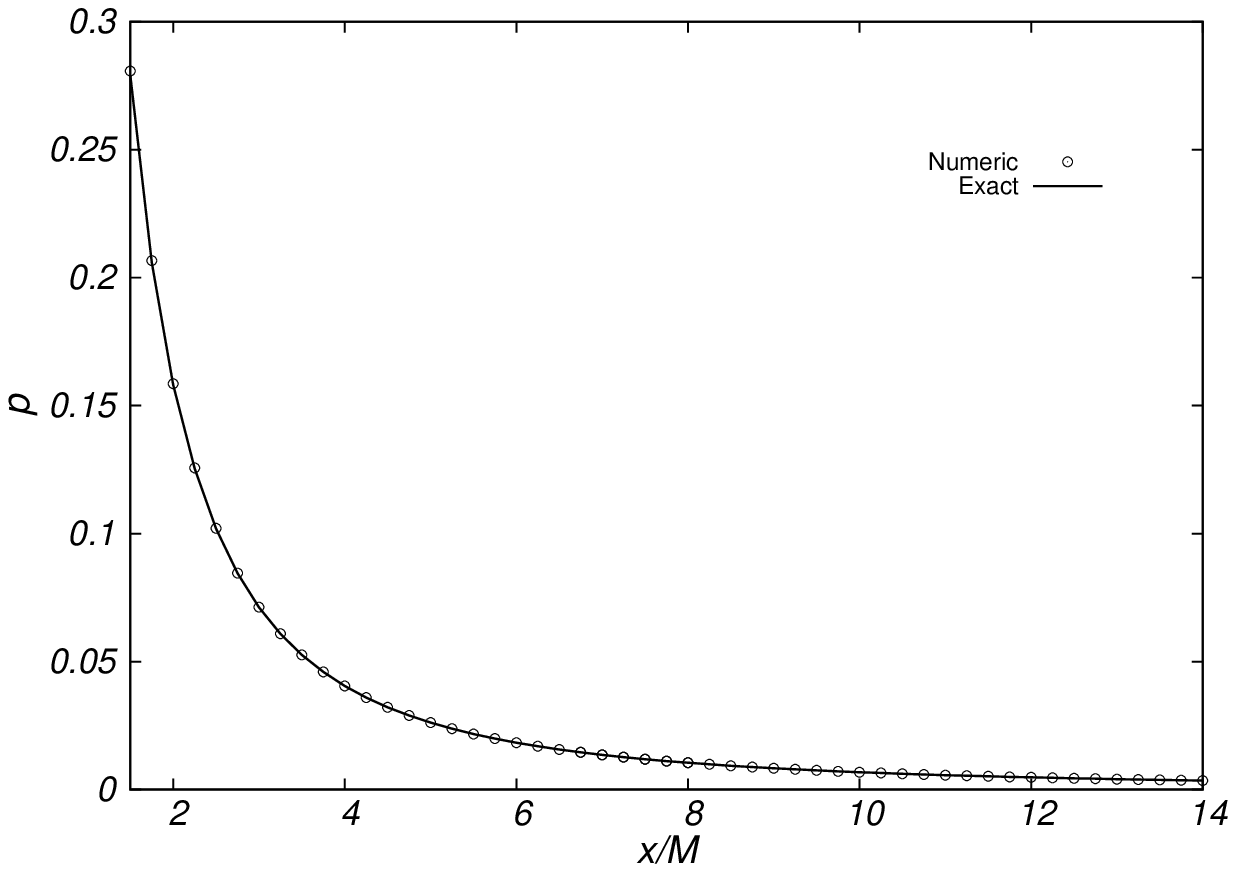}\\
\includegraphics[width=6.5cm,height=6.5cm]{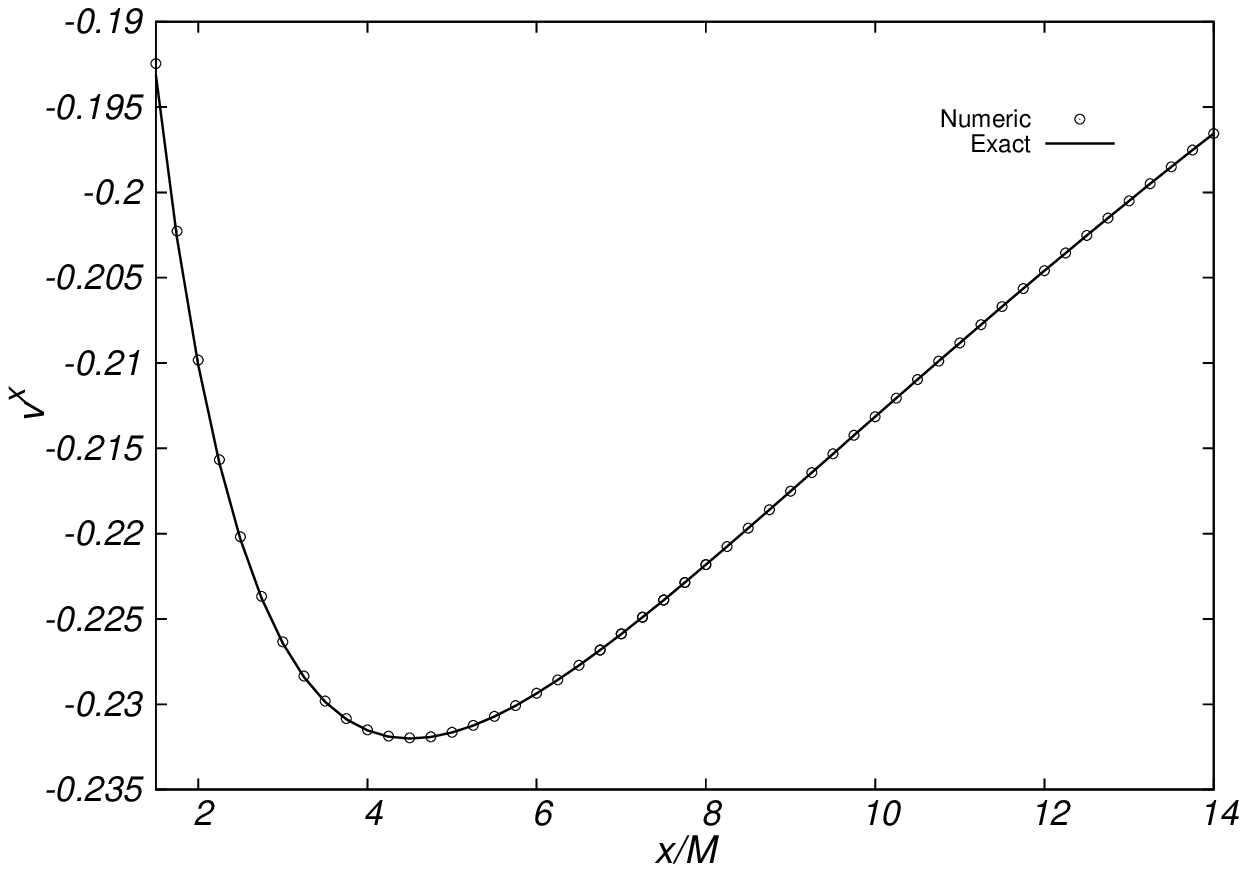}
\includegraphics[width=6.5cm,height=6.5cm]{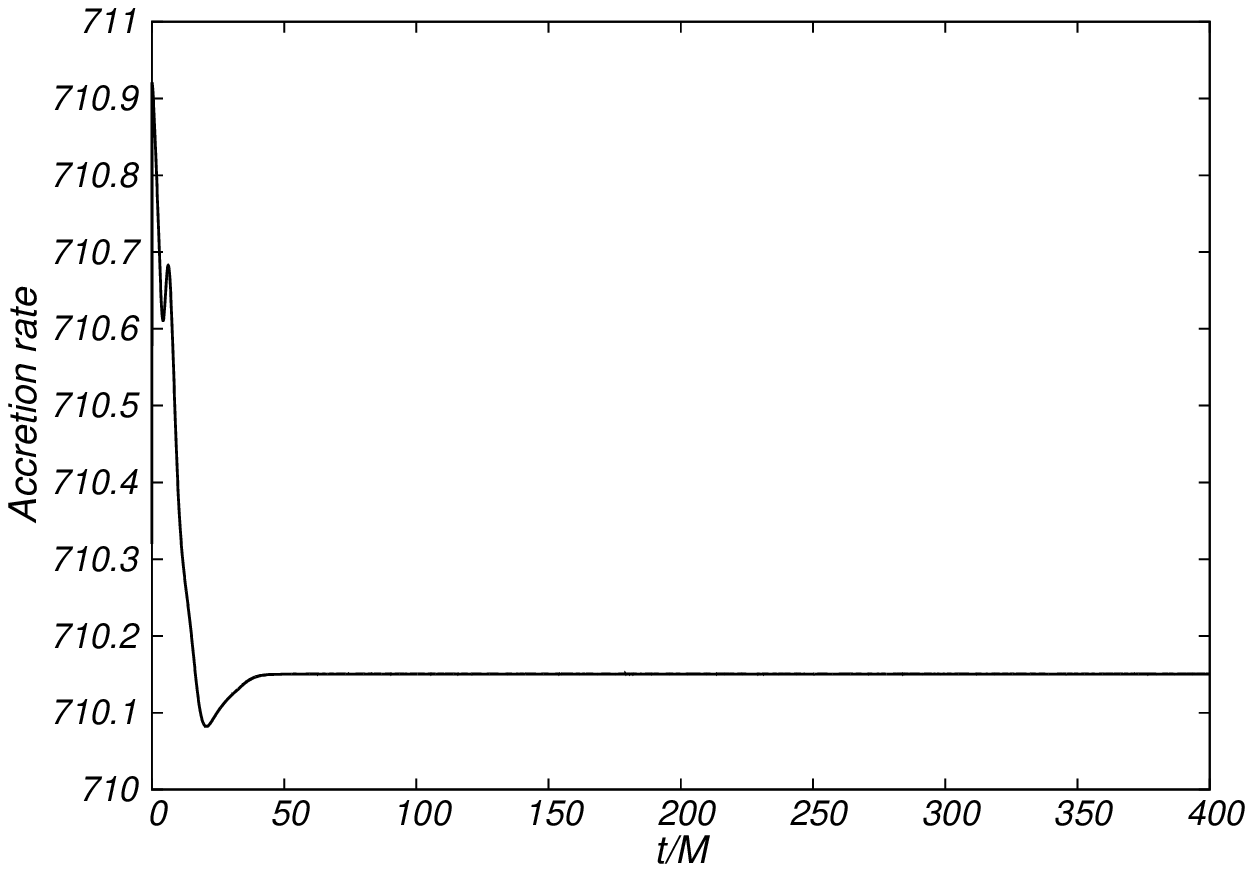}
\caption{\label{fig:michel} Superposed snapshots of the rest mass energy density, pressure and $v^x$ velocity every unit of time from t=0 to $t=400$M along the $x-$axis. We compare the solution with the Michel exact solution we started with. We also show the accretion mass rate $\dot{M}$, which shows an initial transient and afterwards stabilizes.}
  \end{center}
\end{figure}

In order to set the Michel solution as initial condition into the ETK GRHydro thorn, we first interpolate the hydro variables into the ETK 3D cartesian grid, and then perform the tensor coordinate transformation of the velocity field. In order to set a non-rotating black hole space-time we only set the black hole rotation parameter to $a=0$ on the Kerr-Schild metric. Given that Michel accretion is steady, all the hydrodynamical variables have to remain time-independent. We show in Fig. \ref{fig:michel} the superposed snapshots of the evolution of density, pressure and velocity profile projected along the $x$-axis so as the accretion rate measured at a surface located at $r=3$M. The critical values used in this case are $\rho_c = 1\times10^{-2}$, $r_c = 400M$, whereas the polytropic index is  $\Gamma = 4/3$ \cite{papa}. It can be observed that, the density, pressure, velocity profile and the accretion rate remain time independent.

This simple test shows the inflow boundary conditions we implemented are correct, in all the six faces of the boundary. We set the numerical domain to $[15M, 15M ]^3$, covered it using resolution of $0.25M$ in all directions, and set an excision radius to $r_{exc}=1.5$M. We ran the test using HLLE as Riemann solver and the minmod limiter.

\acknowledgments
This research is partly supported by grant CIC-UMSNH--4.9.

\clearpage

\end{document}